\documentclass[a4paper]{article}

\usepackage{amsthm,amsmath}
\usepackage{multirow}
\usepackage{subfigure}

\usepackage{xcolor}
\usepackage{lipsum}                     
\usepackage{xargs}                      
\usepackage[colorinlistoftodos,prependcaption,textsize=small]{todonotes}
 
\newcommand{\pvalueij}{\mbox{$\gamma_{ij}$}}

\providecommand{\keywords}[1]
{
  \small	
  \textbf{\textit{Keywords---}} #1
}

\title{Understanding the growth of the Fediverse through the lens of Mastodon \thanks{Accepted for publication with \textit{Applied Network Science}, Springer Open, June 12, 2021.}}
\author{Lucio {La Cava} \and  Sergio Greco \and Andrea Tagarelli}

\date{\small Dept. Computer Engineering, Modeling, Electronics, and Systems Engineering (DIMES), University of Calabria, 87036 Rende (CS), Italy\\[1ex] \{lucio.lacava, greco, tagarelli\}@dimes.unical.it }

\begin{document}
\maketitle

\begin{abstract} 

Open-source, Decentralized Online Social Networks (DOSNs) are emerging as alternatives to the popular yet centralized and profit-driven platforms like Facebook or Twitter. In DOSNs, users can set up their own server, or instance, while they can actually interact with users of other instances. 
  Moreover, by adopting the same communication protocol, 
DOSNs  become part of a massive social network, namely the \textit{Fediverse}.   
Mastodon is the most relevant platform in the Fediverse to date, and also the one  that has attracted attention from the research community.  Existing studies are however limited to an analysis of a relatively outdated sample of Mastodon focusing on few aspects at  a user level, while several open questions have not been answered yet, especially at the instance level. 
 
In this work, we aim at pushing forward our understanding of the Fediverse by leveraging the primary role of Mastodon therein. 
Our first contribution is the building of an up-to-date and highly representative dataset of Mastodon. Upon this new data, we have defined   a network model over Mastodon instances and  exploited it  to investigate  three major aspects:  
 the structural features of the Mastodon network of instances from a macroscopic as well as a mesoscopic perspective, to unveil the distinguishing traits of the underlying federative  mechanism; 
 the backbone of the network, to discover the essential interrelations between the instances; and 
 the growth of Mastodon, to understand how the shape of the instance network has evolved during the last few years, also when broading the scope to account for instances belonging to other platforms.   
 Our extensive analysis of the above aspects has provided a number of findings that reveal distinguishing features of Mastodon and that can be used as a starting point for the discovery of all the DOSN Fediverse.


\end{abstract}
\hspace{10pt}

\vspace{10mm}
\section{Introduction}

  In the last decade, we witnessed an unprecedented proliferation of Online Social Networks (OSNs). 
  Roughly and generally speaking, OSNs  aim to shrink timing and distances that characterize inter-personal relationships through the Internet. 
  However, the extreme popularity gained by Facebook and the other worldwide available yet  centralized OSN platforms (i.e., hosted and controlled by a single company)  has soon led their owners to pursue a collateral social-marketing goal, which is mostly implemented through content personalization mechanisms and advertisement strategies.  
 As it is well-known, side-effects such as the formation of information bubbles and concerns about the protection of data and user privacy normally characterize most existing centralized OSNs. 

  The above aspects contributed to raise the opportunity for developing new paradigms of OSNs to become   ``user-centric'' rather than ``company-centric'' platforms.  
  As a major consequence, privacy control, as well as  spontaneous and recommendation-free communications among the users, are favored and unbiased  as much as possible from the invasiveness of   advertisements.
  
In this context,  Decentralized Online Social Networks (DOSNs) are emerging as  alternatives to the   popular    centralized platforms.  DOSNs  are built upon two key aspects:  the availability of open-source software to allow everyone to set up their server hence avoiding centralization, and the existence of specific communication protocols to enable fluid interconnections between servers that embrace the same protocol.


These core components lead to a \textit{federation} model, in which  the servers, also called as \textit{instances}, can communicate to each other  through the same protocol. This implies that users which are signed up for a particular server can actually interact with users of other servers, analogously to what normally  happens  with email services. 
DOSNs hence   become part of a massive social network, namely the \textit{Fediverse}.  As a consequence of this mechanism, users can use their accounts on a DOSN platform to follow users on other platforms, without needing an account there. 

The Fediverse currently provides several services, such as \textit{Mastodon} and \textit{Friendica} for microblogging,  \textit{PeerTube} and \textit{Funkwhale} for video hosting,  \textit{PixelFed} for image hosting. Among these platforms, Mastodon is the one that has encountered the greatest attention increase over the years. 
Mastodon provides a user experience comparable to Twitter (e.g., 
published contents are called \textit{toots}, whereas the analogous of the retweet functionality is  called \textit{boost}) and Reddit (e.g., niche communities and content moderation are emphasized, however Mastodon communities are independent of each other). Moreover, Mastodon affords content warnings, i.e., synopses of toots that can preview disturbing content. 
Mastodon is built upon the \textit{ActivityPub} protocol,\footnote{https://www.w3.org/TR/activitypub/}   which implements a layer for client-to-server communications and another one for the server-to-server communications. Thanks to this protocol and a subscription-based mechanism  (implicitly carried out by the instances), users can interact with each other even if they  belong to different instances. 

The extended followship mechanism in Mastodon also leads to an original yet remarkable timeline structure, namely  \textit{home}-timeline, which   provides toots generated by followed users,  \textit{local}-timeline, which  yields toots created within the instance, and  \textit{federated}-timeline, which contains all public toots from all users (either from the same instance or not) that are known to the instance where a user is registered.

Furthermore, Mastodon instances allow their users to apply  rules and policies on the generated contents. Administrators can declare both the main topics of their instance and prohibited contents. Users can mark some contents as inappropriate for a given instance by placing a \textit{content warning}   on the content itself. Along with the content warning, a spoiler (i.e., a textual component summarizing the obfuscated content) will be displayed   to the user, letting her/him decide whether to view it or not.

Finally, Mastodon also allows administrators to close registrations for their instances, e.g., in the case of a ``private instance'' among friends, or to efficiently moderate contents. Nevertheless, this feature does not affect user  interactions which, as outlined above, are guaranteed by specific protocols.

\vspace{2mm}
\noindent 
\textbf{Related Work.}  
 DOSN analysis is a relatively novel research field.  
Early works mainly investigated motivations, opportunities and challenges related to different  solutions for the decentralized paradigm, 
 from  distributed systems like peer-to-peer networks to hybrid systems integrating external and private resources for storing user data~\cite{GuidiCPR18}. Two surveys on these topics as   well as on issues related to DOSN infrastructures, data management, privacy and information diffusion, can be found in~\cite{GuidiCPR18, datta2010}.   
 
Focusing on more recently developed and open-source DOSNs, to the best of our knowledge,  Mastodon is the only platform in the Fediverse that has received noticeable attention from the research community~\cite{CerisaraJOL18,abs-1811-09292,Zignani2018,Raman2019,Zignani2019,Zulli2020}.  
Zulli et al.~\cite{Zulli2020} have recently performed a qualitative analysis based on an interview to a sample of instance moderators. From that  study it emerges that  
 the federative structure of Mastodon enables content variety and community autonomy, and also emphasizes horizontal growth between instances rather than growth within instances;  
 however,  any analysis of the interactions on the Mastodon instances is   missing.
 
 From a network science perspective,  the studies by Zignani et al.~\cite{Zignani2018, Zignani2019} are particularly relevant, as they were the first to analyze a portion of the Mastodon user-network,  focusing on degree distribution,  triadic closure,  and assortativity aspects, and comparing such characteristics to those in Twitter~\cite{Zignani2018}.  
 From the analysis of the in-degree and out-degree distributions, 
 Mastodon is found to show a more balanced behavior   between followers and followees than what observed in Twitter.  
Also, the 95\% of users exhibit a difference between followers and followees bounded in the range (-250, 250).  Concerning social bots,  the authors  reported a low presence (around 5\%), which is significantly lower than the 15\% observed on Twitter by Varol et al.~\cite{Varol2017}.  Clustering coefficient in Mastodon ranges between those of   Facebook  and Twitter. 
The degree assortativity in Mastodon was also inspected,      considering  source in-degree  (SID),  source out-degree  (SOD),  destination in-degree  (DID), and  destination out-degree  (DOD).  The authors observed lack of correlation between (SOD, DOD),  (SOD, DID) and (SID, DOD),  which indicates that 
users who follow many users are connected to users whose popularity may vary (DID) and who in turn follow few or many users (DOD).  Moreover,  the observed negative correlation (-0.1) between SID and DID implies that the higher the popularity of a user is,  the less popular the users s/he follows will be.  Overall,   disagreement is observed between the degree assortativity in Mastodon and the ones shown by well-known social networks.  Finally,  Zignani et al.  found that users' hubiness is bounded within its instance and influenced by the latter.  
 Also,  in~\cite{Zignani2019},  the authors investigate how the decentralization process affects relationships between users,  unveiling that instances show individual footprints (based on degree distribution and clustering coefficient statistics observed on the top 10 instances in Mastodon) that influence relationships. 

It should be noted that none of the above works have focused on the interactions between instances in Mastodon, and a number of related aspects are missing, such as distinction between online and offline instances,   mesoscopic structure analysis, or backboning.  
 Indeed, several open questions still remain to address on Mastodon,  which inspired our study in this work. 


\vspace{2mm}
\noindent 
\textbf{Contributions.} 
Our research stems from a twofold motivation: to provide a fresh view on Mastodon based on recently updated data, 
 and to fill a lack of knowledge on  topological  features of the Mastodon network focusing at the \textit{instance} level.\footnote{An abridged version of this work appeared in the \textit{Proc. of the 9th Int. Conf. on Complex Networks and their Applications, 2020}~\cite{LaCavaRT20}.} 

 As previously discussed, early studies have primarily focused on the analysis of Mastodon users, and they captured a relatively small snapshot of Mastodon dated four years ago.  
 Clearly, this might have overlooked salient traits that can be discovered  at the  instance  level, as well as it raises the need for getting a timely picture of Mastodon which has presumably changed over time. 
To overcome these limitations, our study builds upon an up-to-date and representative network data over the instances, and utilizes it to provide insights into their relations. 
The  goal is manifold:  it includes the opportunity of   enhancing our  understanding of the \textit{macroscopic} and \textit{mesoscopic} structures of Mastodon  to unveil the distinguishing traits of the underlying federative  mechanism, and to discover the essential interrelations between the instances; but also we want to understand how the instance network has changed, within Mastodon as well as at the boundary of Mastodon itself.

We elaborate on the above aspects by developing an extensive analysis framework to answer the following research questions:

\begin{enumerate} 
\item[\textbf{Q1}] -- \textit{Network data and models}:   How are the Mastodon instances detected and modeled as a network? 
\item[\textbf{Q2}] -- \textit{Structural features}:  What are the salient structural features of the network of Mastodon instances, at \textit{macroscopic} as well as \textit{mesoscopic} level?
\item[\textbf{Q3}] -- \textit{Fingerprint}:  Are there any clues to the presence of notable phenomena that distinguishes Mastodon from centralized OSNs? How does 
a federative mechanism arise from  the Mastodon instances? 
\item[\textbf{Q4}] -- \textit{Network backbone}:  What is the backbone of the network of Mastodon instances, and does it preserve the  structural features of the whole network? 
\item[\textbf{Q5}] -- \textit{Growth}:  How has the shape of the network of Mastodon instances  evolved during the last few   years?  
\end{enumerate}
 
%

\vspace{2mm}
\noindent
{\bf Plan of the paper.\ }
 The remainder of the paper is organized so as to address the above stated  research questions. 
 Section~\ref{sec:data} describes our crawling methodology, the data collected and the network models we built upon this data (\textbf{Q1}). 
 Section~\ref{sec:structural} contains the structural analysis of the Mastodon instance network, from the macroscopic and mesoscopic perspectives (\textbf{Q2}-\textbf{Q3}). 
 Section~\ref{sec:backboning} describes our methodology of identification of the backbone of the Mastodon instance network (\textbf{Q4}). 
 Section~\ref{sec:evolution} analyzes the evolution of the Mastodon instance network  (\textbf{Q5}) from three points of view: comparison with the earlier Mastodon network, emphasis on the online portion of the network, and an analysis of  centrality of the instances. 
 Finally, Section~\ref{sec:conclusion} concludes the paper and provides pointers for future research.

\section{Polite data crawling and network modeling}
\label{sec:data}

  To answer our first research question (\textbf{Q1}), here we describe the crawling methodology adopted to collect public information from Mastodon, the steps carried out to build and validate our  Mastodon instance dataset, and the network models we derived from the collected data.

\subsection{Crawling methodology} 
The publicly available dataset on Mastodon relationships provided in~\cite{Zignani2018} contains data extracted during the period  between 2017 and 2018. 
This clearly raises concerns about the possibly partial obsolescence of those data, since  social networks continuously evolve and there is no reason to assume that Mastodon and the Fediverse would represent an exception to this rule. 
Therefore, to satisfy the need for up-to-date data, we carried out an extensive crawling phase based on a newly designed crawler.

\vspace{1mm}
{\bf Crawling requirements and design principles.\ }  
Our crawler was developed under strict and self-imposed constraints,  i.e., following the \textit{privacy by design, privacy by default} approach, and exclusively relying on the publicly-available Mastodon REST APIs~\footnote{https://docs.joinmastodon.org/api/} --- using such APIs, we accessed data through \textit{GET} and \textit{POST} methods of the  HTTP  protocol, and managed the payload of requested data in a  JSON format.
 Under these constraints, we were able to make our crawling methodology fully compliant with ethical and privacy-related principles.

Given the decentralized nature of Mastodon, it is not straightforward to  detect the myriad of instances available today. 
Nonetheless, to get updated information on the current landscape of Mastodon instances, the  \textsl{instances.social} website~\footnote{https://instances.social/}  is commonly used as a de-facto tracker of Mastodon instances. 
We exploited it  to generate a list of seeds (i.e., starting points for the searching process), which correspond to  currently online Mastodon instances.  

 Mastodon instances provide developers with \textit{authentication tokens} to ensure control over the scope of the interactions. Moreover, by leveraging on authenticated requests, developers might achieve better interaction capabilities with instances. These conditions certainly comply with our desired privacy and ethical principles. Therefore, we submitted our seed list (i.e., the   instances obtained from \textsl{instances.social}) to the authentication process, getting approved from approximately 900 instances out of about 1\,100. Also, being able to traverse instances timeline --- via authenticated requests ---  we discovered about 81\,000 new users to explore. Then, we carried out a \textit{breadth-first-search} over them, detecting incoming and outgoing links and progressively increasing the number of users to explore, by discovering new ones during the link detection process. 
 
We point out that Mastodon allows redirecting or moving a user's profile. Although notable, this feature could determine some inconsistencies during the crawling, such as redirects to other instances while exploring the user profiles. Therefore, we avoided generating edges for users who presented similar behaviors.
 
Moreover, two side yet relevant remarks arise regarding our crawler implementation. First, to efficiently handle the  collected data, we used a caching mechanism (\textit{Redis}) coupled with a   NoSQL database (\textit{MongoDB}).  
Also, to prevent computational bottlenecks, we avoid repeated checkings over the database during the crawling phase (e.g., for checking duplicated edges), so that we eventually refine the complete network dataset in an ``off-line mode'', once the crawling process has finished, by exploiting  particularly efficient processing functionalities  provided by suitable  data and network manipulation software libraries. 

 
 Our crawling session ended up with 27\,989\,557 
  links detected.  
  After performing basic  data-cleaning steps, particularly removing duplicate links, we obtained about 1.4M and 18M unique users and links, respectively, managing to cover 16\,282 instances.
  
It should be emphasized that, to respect privacy principles, we firmly avoided using scraping techniques or systems, i.e., we abstained crawling information from instances which did not provide us with an authentication token. 
Notice also that the detected links were immediately anonymized, and any information that could impact on the users’ privacy was  replaced with numerical data generated through a proper hashing function --- as a consequence, it is not possible   to trace back the original information on users  from our raw dataset.
Finally, we point out that our fetching of descriptive text data (e.g., toots) was \textit{minimal}, i.e., it occurred only during the initialization  of our crawling process: indeed, we produced the seed-user set by discovering them through toots available in the timelines of the seed instances, relying only on authenticated requests. Nonetheless,  we never stored this data since we processed it in real-time. After this initial phase, the crawling continued  via  breadth-first-search, thus  ignoring   textual data.

\vspace{1mm}
{\bf Spotting Mastodon instances.\ }
  As previously mentioned, platforms in the Fediverse utilize a shared protocol, allowing for seamless interactions among their users. A related key-aspect is that, when requesting followings or followers of a Mastodon user, the APIs return all of them, regardless of the Fediverse platform.  
  In this regard, one question becomes how we can distinguish between instances that belong to Mastodon from other platforms' instances in the Fediverse. We answered this question through a verification process, as summarized next.

\begin{table}[t!]
\centering
\caption{ Current landscape of Mastodon instances as provided by \textsl{instances.social} and \textsl{fediverse.party} websites. Symbol $\bigcup$, resp.   $\bigcap$, stands for the total of Mastodon instances calculated as the size of the set union, resp. intersection, between the instances set provided by the websites.}
\label{tab:stats-instances}
\vspace{2mm}
\rmfamily
\begin{tabular}{|l||c|c||c|c|}
\hline 
& \textsl{instances.social} & \textsl{fediverse.party} & $\bigcup$ & $\bigcap$ \\
\hline 	\hline
Online & 1\,193  &  not available & 1\,193 & 997 \\  
Online+Offline & 7\,313 &  3\,396 & 9\,433 & 1\,276 \\
\hline
\end{tabular} 
\end{table}

To date, some relevant websites provide up-to-date Mastodon information, namely the aforementioned \textsl{instances.social} and \textsl{fediverse.party},\footnote{https://fediverse.party/en/mastodon}   so that we exploited them to filter our data through their lists of known instances.  
Note however that, while \textsl{fediverse.party} does not  distinguish   online from offline instances,   \textsl{instances.social} provides fine-grained filtering capabilities. 
 In this regard, we focused on the setting of two main parameters provided in the \textsl{instances.social} APIs: \textit{include\_dead} and \textit{include\_down}. As declared in \textsl{instances.social}, an instance is considered \textit{dead} if inactive for at least two weeks, and \textit{down} if it is not currently online yet live within a two-week window. We set either both \textit{include\_dead} and \textit{include\_down} to \textit{true}, or \textit{false}, to obtain all the known Mastodon instances, resp.  online-only instances.

As reported in Table~\ref{tab:stats-instances}, we merged information retrieved from both websites, regardless of the instances' status (i.e., online or offline), obtaining 9\,433 known Mastodon instances. In addition, we requested online-only ones to \textsl{instances.social}, getting 1\,193 instances to date. 
Note also that, by restricting our census of the Mastodon instances to the information shared between the two platforms, the total online and the overall total would be decreased of 16.4\% (997)  and 86.5\% (1\,276).

\begin{figure}[t!]
\centering
\includegraphics[width=0.7\textwidth]{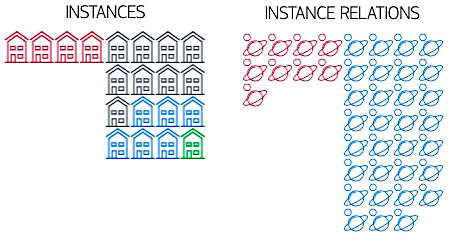}\\ \vspace{2mm}
\rmfamily
\scalebox{0.8}{
\begin{tabular}{|l||c|c|c|c|}
\hline
& \textit{Total collected} & \textit{Mastodon Overall} & \textit{Mastodon Online} & \textit{Time} \\
\hline 	\hline
This work & 16\,282  &  6\,960 & 1\,116 & Late 2020 \\ 
Earlier~\cite{Zignani2018}  & not available & 4\,015 & 548 & Mid 2017 - Early 2018 \\
\hline 
\end{tabular}
} \\ \vspace{2mm}
\caption{
Validated data based on the  information reported in Table~\ref{tab:stats-instances}, and illustrative comparison between dimensions of the  earlier state-of-the-art (in red) and dimensions of our dataset. 
}
\label{fig:dataset-size-comparison}
\end{figure}

\vspace{1mm}
\textbf{Validated datasets.\ }
Based on the above information,  we analyzed our crawled data to properly detect the status of the instances. 
Results are summarized in Figure~\ref{fig:dataset-size-comparison}. 

We intercepted 
 6\,960 out of 9\,433 Mastodon instances (both online and offline), and 1\,116 out of 1\,193 currently online instances. We point out the significance of the latter value, given the coverage of most of the online Mastodon instances to date. Moreover,  our dataset doubles the earlier state-of-the-art in terms of currently online instances. Clearly, the freshness of our data (November-December 2020) influences this value.

It should be emphasized that  our collected data includes a remarkable amount (9\,322) of non-Mastodon instances, i.e., belonging to other Fediverse platforms. This clearly  strengthens the concept of Fediverse, 
but also opens to the discovery of the  primary role taken by Mastodon within the Fediverse. 
 In fact, although  we detected non-Mastodon users and instances through Mastodon ones, and hence this knowledge of the Fediverse might be partial, our collected data offers an unprecedented opportunity for deepening our understanding of the position of Mastodon in the Fediverse (i.e., how Mastodon instances and users interact with the rest of the Fediverse), 
 given the premises of independence yet cooperation among platforms in the Fediverse. 


Further important remarks also arise  regarding the growth of Mastodon. Although we are not aware of the status (i.e., online or offline) of the instances in the earlier state-of-the-art dataset~\cite{Zignani2018} at the time of their  creation, 
we hypothesize that, after a first boost due to enthusiasm and novelty,  Mastodon reached its stability as a DOSN. Indeed, the number of currently online instances is moderate compared to the number of all-time known ones and refers to non-transient yet well-rooted platforms.  
Overall, our dataset turns out to be significantly larger and more recent   than the earlier Mastodon dataset, making it more suitable for novel and further studies.  

\subsection{Network models}
\label{sec:models}


Let us denote with $\mathcal{U}$  the set  of users and with   $\mathcal{I}$ the set of instances available in the   extracted Mastodon data.  
We can define a directed network modeling the Mastodon data   as $\mathcal{G} = \langle \mathcal{V}, \mathcal{E} \rangle $,  where the node set $\mathcal{V}$ contains pairs $(u,i)$,  with $u \in \mathcal{U}$ and $i \in \mathcal{I}$,  and the edge set $\mathcal{E} \subseteq \mathcal{V} \times \mathcal{V}$ corresponds to the set of following relations,   such that any $(x, y) \in \mathcal{E}$ with $x=(u,i)$ and $y=(v,j)$  means that user $u$ in instance $i$ follows user $v$ in instance $j$.  
Note that  $u$ may concide with $v$ provided that  $i\neq j$. 
 Given $\mathcal{G}$,  we derive three  Mastodon networks   at   instance  level, which are  formally defined as follows.

 \textit{Instance network.\ }  
The first network we define is the  graph modeling relations between all the instances in $\mathcal{I}$, hereinafter referred to as \textsc{Instances} network,  as the directed weighted network $G_{\mathcal{I}}=\langle V, E, w \rangle$,  where $V=\mathcal{I}$ is the set of nodes, $E$ is the set of edges such that $(i,j) \in E$ means that there exists at least one   user in instance $i$ that follows another user in instance $j$,  and $w:E \mapsto \mathcal{R}$ is an edge weighting function such that,  for any $(i,j) \in E$, $w(i,j)$ stores the multiplicity of the following relation from $i$ to $j$ (i.e.,  number of users in $i$ following users in $j$). 

 \textit{Online instance network.\ }  Our second network is induced from the set of instances that are detected as online at the time of the crawling process we carried out. Therefore, by denoting with $V^{o} \subseteq \mathcal{I}$ the set of online instances, the \textsc{Online-Instances} network   $G^{o}_{\mathcal{I}}=\langle V^{o}, E^{o}, w^{o} \rangle$, with edge-set $E^{o} =  E \cap (V^{o} \times V^{o})$ and edge weighting function $w^{o}:E^{o} \mapsto \mathcal{R}$, is defined to model  the connections between the online instances only. 


\begin{table}[t!]
\centering
\caption{Networks created from our collected dataset, and comparison with the earlier state-of-the-art network.  All networks but \textsc{Expanded} refer to Mastodon-only instances.}
\label{tab:stats-network}
\vspace{2mm}
\rmfamily
\begin{tabular}{|l||c|c|}
\hline
 Network name & \textit{\#Nodes} & \textit{\#Edges} \\
\hline 	\hline
\textsc{Expanded-Instances} & 16\,282  &  318\,218  \\ 
\textsc{Instances}  & 6\,960 & 216\,504 \\
\textsc{Online-Instances} & 1\,115 & 75\,046 \\
{Earlier}~\cite{Zignani2018} & 4\,015 & 95\,221 \\
\hline
\end{tabular}
\end{table}



   \textit{Expanded network.\ } Our third network generalizes the first one by   accounting for  instances that have been recognized   outside Mastodon. Actually, every link extracted during our crawling process is by definition incident with at least one instance that belongs to Mastodon. 
Therefore, we also define an  expanded network  to explore the boundary of the Mastodon network to the rest of the Fediverse. 
By denoting with   $V^* \supset \mathcal{I}$ such  expanded set of instances, i.e., the whole set of crawled instances, we  define the \textsc{Expanded-Instances} network  as $\mathcal{G}^*_{\mathcal{I}} = \langle V^*, E^*, w^* \rangle$, where 
 $E^*= E \cup \{(i, j) \ | \ (i \in V \land j \in V^* \setminus V) \lor (i \in V^* \setminus V \land j \in V)\}$, 
and the weighting function $w^*:E^* \mapsto \mathcal{R}$ follows analogous definition as for the Mastodon instances  network. 

\begin{figure}[t!]
\centering
\includegraphics[width=0.97\textwidth]{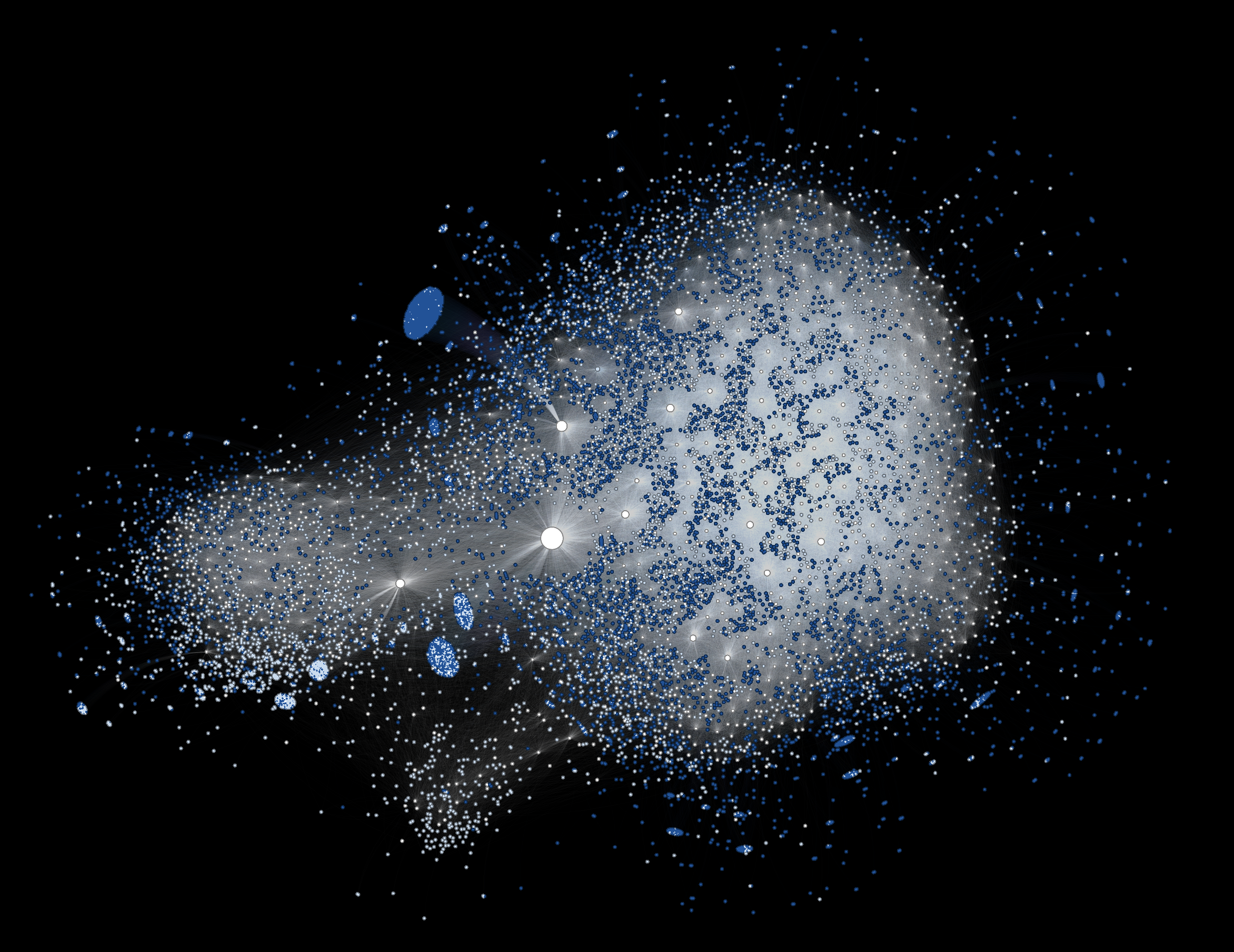} 
\caption{
Illustration of the \textsc{Expanded-Instances} network.  Node colors denote different types of instances: white and light blue indicate online and offline Mastodon instances, respectively, whereas dark blue corresponds to non-Mastodon instances. Node size is proportional to its degree. The displayed layout is based on the force-directed drawing ForceAtlas2 model. (\textit{Produced by using the Graphistry service, available at https://www.graphistry.com/}.)
}
\label{fig:plot}
\end{figure}

All the above defined networks and the one inferred from the earlier dataset are summarized in Table~\ref{tab:stats-network}.  
Note that the number of nodes in the \textsc{Online-Instances} network is decreased of one w.r.t. the information given in Fig.~\ref{fig:dataset-size-comparison}, since one online instance is actually disconnected from the network.  
 Moreover, in Figure~\ref{fig:plot}, we provide an illustration of the overall network of instances created from our collected dataset, i.e., \textsc{Expanded-Instances}.  

Remarkably, the \textit{Earlier} network is found to be mostly    contained in our \textsc{Instances} network; more precisely, about 80\% of the instances in the \textit{Earlier} network are also contained in our \textsc{Instances}. 
 As we shall discuss later in this work, this has important implications in the growth of Mastodon during the last three years.  

It should also be noted that our collected data allows us in principle to build networks at the \textit{user} level as well, e.g., we could define the network     of the relations between users of  each particular instance; 
 nonetheless, this goes beyond the scope of this work, whose focus is the analysis and understanding of the relations among the instances in Mastodon. Therefore, we leave the study of instance-specific networks of users as future work (cf. Section~\ref{sec:conclusion}).

%

\section{Structural analysis of the {\rmfamily\textsc{Instances}} network}
\label{sec:structural}



In this section, we answer our second research question (\textbf{Q2}) by presenting an extensive analysis  of the network we built over the Mastodon istances, i.e., the previously introduced  \textsc{Instances} network. 
To unveil the main characteristics that define Mastodon, we will take a  
 macroscopic as well as a  mesoscopic perspective, and organize the discussion into  the two next subsections. 


\begin{table}[t!]
\centering
\caption{Summary of structural characteristics of the \textsc{Instances} network, including details on community structure and core decomposition.}
\label{tab:fullstats}
\vspace{2mm}
\rmfamily
\scalebox{0.9}{
\begin{tabular}{|l||c||c|c|c|}
\hline
& \textsc{Instances} & \multicolumn{3}{c|}{\textsc{Instances}  inner-most core}\\
\cline{3-5}
&   & \textit{degree} & \textit{in-degree} & \textit{out-degree} \\
\hline 	\hline
\#nodes & 6\,960  & 189 & 208 &  196 \\ 
\#edges & 216\,504 & 25\,790  & 28\,690 & 26\,463  \\
reciprocity   &   65.1\%  &   88.4\%  &  85.7\%  &  88.2\%  \\
density   &   0.004  &   0.726  & 0.666  & 0.692  \\
average degree$^*$ &  41.966  & 152.328  &  157.702  & 150.98 \\
average in-degree   &  31.107  &  136.455  &  137.933 &  135.015 \\
\% sources  &   12\%   &  0\%   &  0\%   &  0\% \\
\% sinks   & 6.6\%  &   0\%  &   0.005\% &  0\%  \\
degree assortativity$^*$ & -0.274 & -0.117  & -0.158 & -0.135  \\
degree assortativity & -0.253 & -0.14 & -0.171 & -0.151 \\
average path length   &  2.330  &  1.270  &  1.330  & 1.310  \\
diameter   &   5  &  2  &  2  &  2  \\
transitivity$^*$   &   0.128 &   0.832  & 0.798 &  0.807  \\
clustering coefficient$^*$   & 0.836  &  0.837  &   0.810  &  0.816  \\
clustering coefficient {\scriptsize \textit{(full averaging)}}$^*$   &  {0.687}    &   {0.837}    &   {0.810}    & {0.816}   \\
\#strongly connected components    &   1\,305  &   1  & 2  & 1  \\
\#weakly connected components$^*$   &   1   &  1   &  1   &  1  \\
\hline \hline 
modularity {\scriptsize by \textit{Louvain}}$^*$ &  0.289  & 0.032 & 0.039 & 0.037 \\ 
\#communities {\scriptsize by \textit{Louvain}}$^*$ &  5 (5) & 3 (3) & 3 (3) & 3 (3)\\ 

modularity {\scriptsize by \textit{Louvain}}$^{**}$ & {0.353} &  {0.242} &  {0.246} & {0.246} \\ 
\#communities {\scriptsize by \textit{Louvain}}$^{**}$ & 6 (8) & 4 (5) & 3 (4) & 4 (6)\\ 

\#communities {\scriptsize by \textit{Infomap}}$^{**}$ & 6 (54) & 1 (3) & 1 (4) & 1 (3)\\ 

\hline
\end{tabular} 
}

\vspace{1mm}
{\footnotesize
\begin{flushleft} $^*$ Statistic calculated by discarding the edge orientation 
\end{flushleft}
}
{\footnotesize
\begin{flushleft} $^{**}$ Statistic calculated by taking into account the edge weights 
\end{flushleft}
}
\end{table}

\subsection{Macroscopic structural analysis}
\label{sec:macroscopic}

We begin our investigation of the    \textsc{Instances} network at a  macroscopic level.  
We refer to Table~\ref{tab:fullstats} for a summary of statistics on the main structural characteristics of the \textsc{Instances} network, each of which is analyzed in the following.

\begin{figure}[t!]
\centering
\includegraphics[width=0.6\textwidth]
{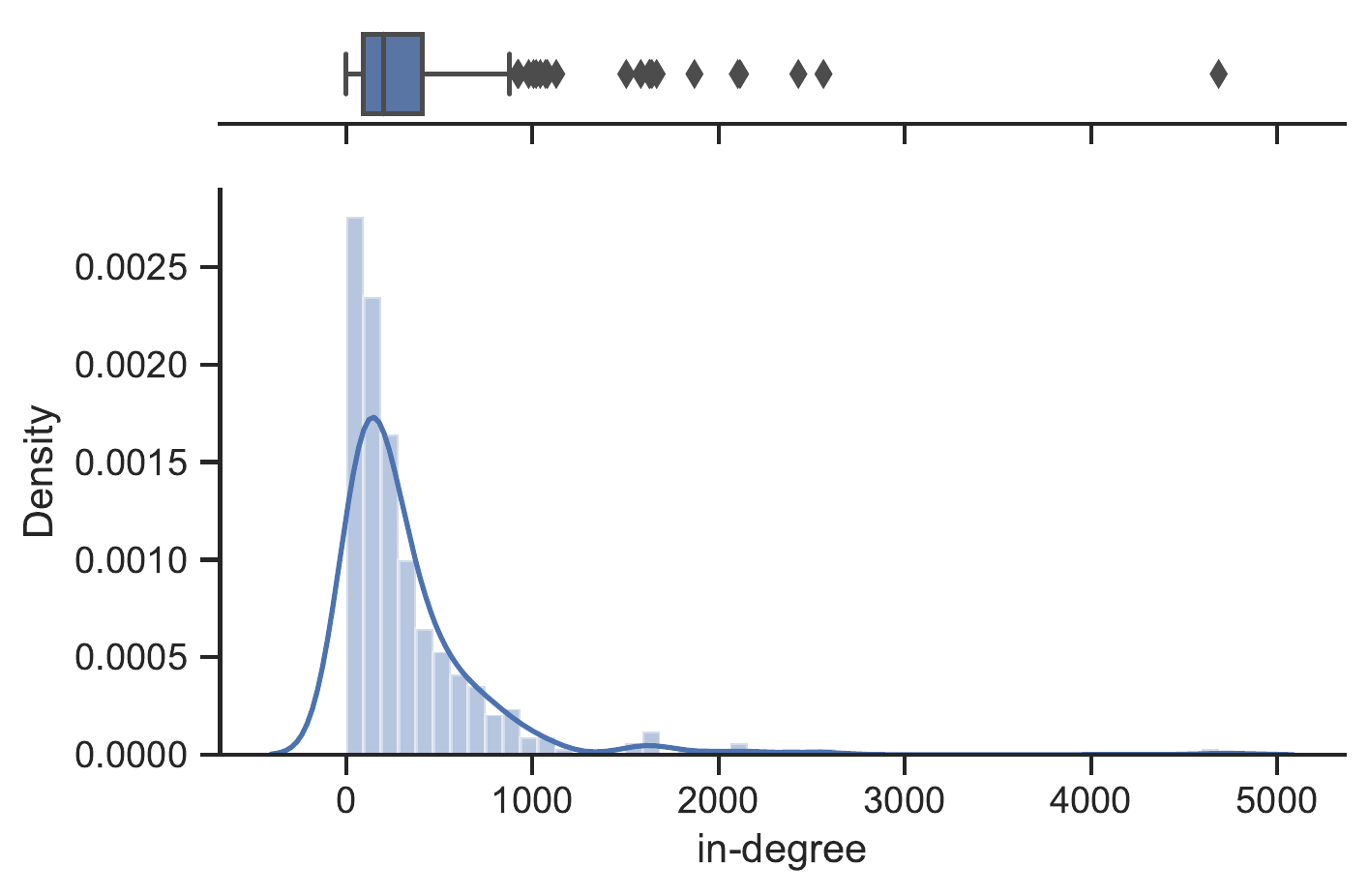} 
\\
\includegraphics[width=0.57\textwidth]
{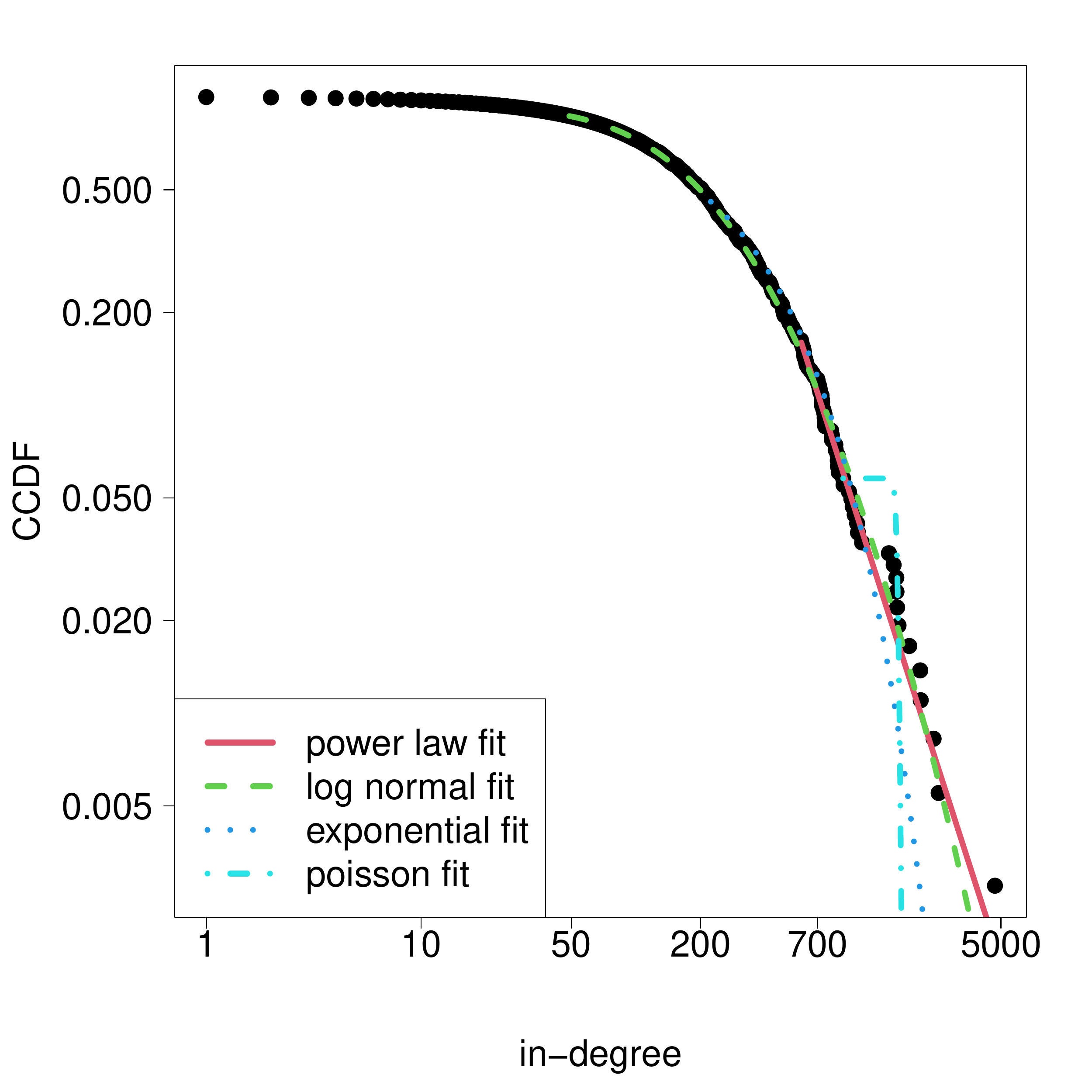} 

\caption{\textsc{Instances}   in-degree distribution: boxplot and Probability Density Function  (top), and Complementary Cumulative Distribution Function, with various distribution fittings (bottom).}
\label{fig:mastodon-instances-fitting-in}
\end{figure}

\begin{figure}[t!]
\centering
\includegraphics[width=0.85\textwidth]
{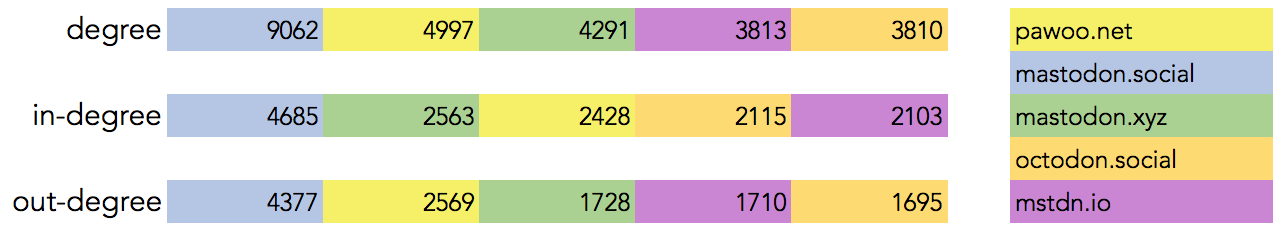} 
\caption{Top-5 instances by degree, in-degree, and out-degree, respectively, in the \textsc{Instances} network.}
\label{fig:top5}
\end{figure}

\begin{figure}[h!]
\centering
\includegraphics[width=0.6\textwidth]
{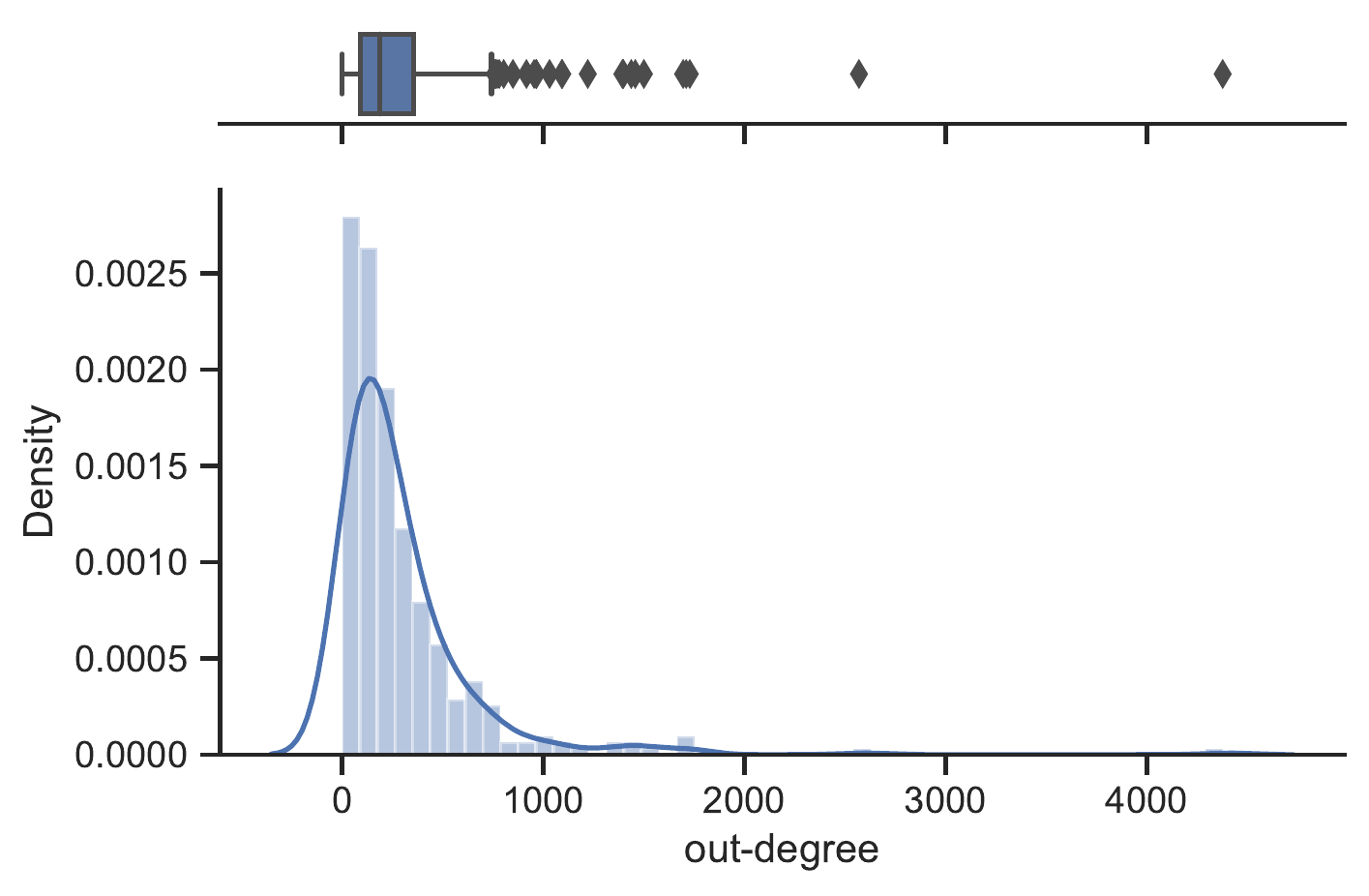} 
\\
\includegraphics[width=0.57\textwidth]
{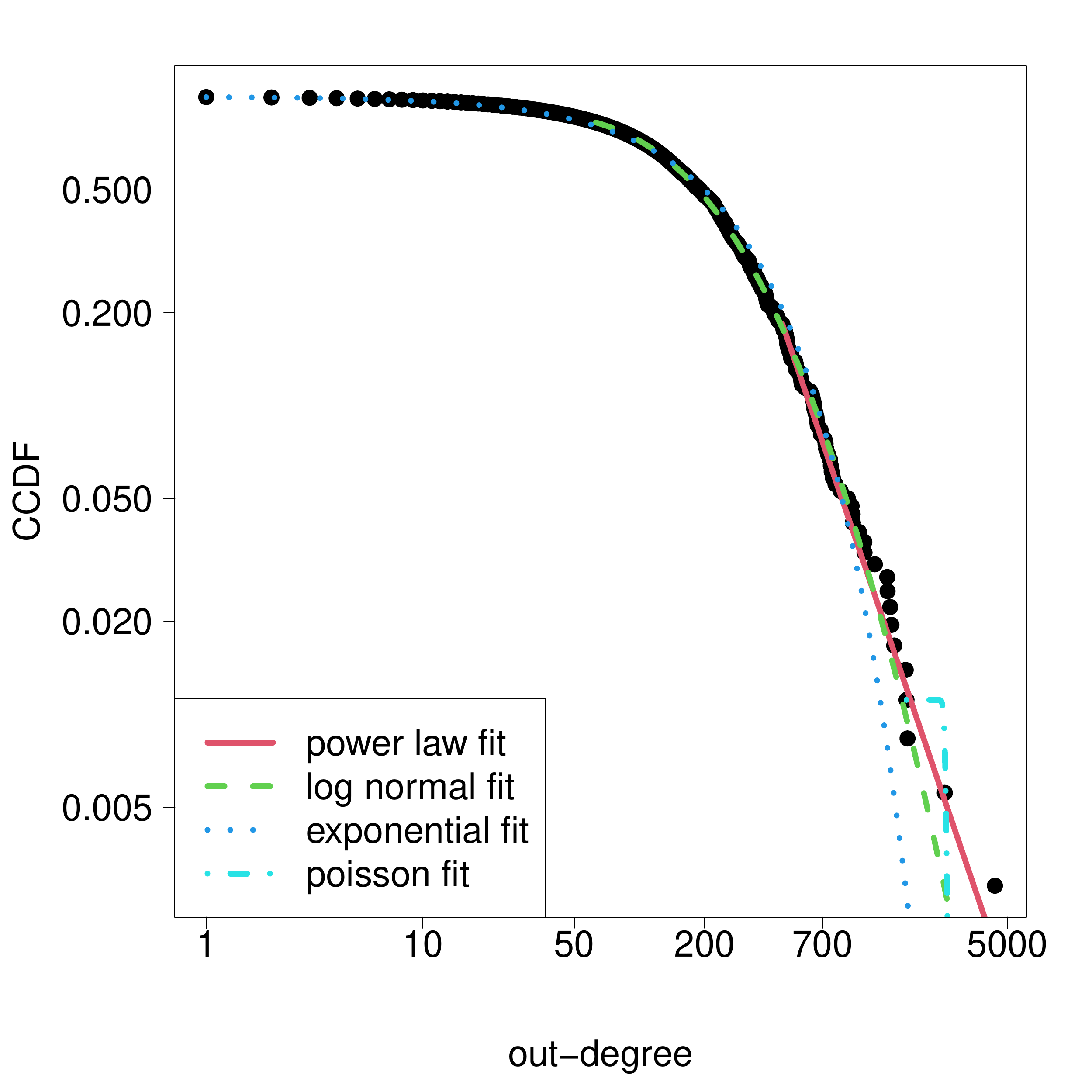} 

\caption{\textsc{Instances}   out-degree distribution: boxplot and Probability Density Function  (top), and Complementary Cumulative Distribution Function, with various distribution fittings (bottom).}
\label{fig:mastodon-instances-fitting-out}
\end{figure}

\textbf{Degree distribution.\ } 
Figure~\ref{fig:mastodon-instances-fitting-in} shows the boxplot, density function, and Complementary Cumulative Distribution Function (CCDF) of the \textsc{Instances} network in-degrees, with various types of distribution fittings;  results   obtained for the out-degrees and the total degrees are analogous, and  we report them in the Appendix (Figs.~\ref{fig:mastodon-instances-fitting-all}--\ref{fig:mastodon-instances-fitting-out}). 

Looking at the boxplot and the histogram with associated density function,  there is evidence of right-skewness of the in-degree distribution, with a small bunch of   ``outliers'' scaling in the regime of thousands. 
In detail, while the first quartile, median, mean, third quartile, and non-outlier maximum degree are 91.5, 202, 331, 413, and 878, respectively, there are 21 instances having in-degree above 880, up to a maximum degree of  4\,685.  

We investigated about the outlier instances, focusing on the top-5 by degree, resp. in-degree and out-degree, as shown in Figure~\ref{fig:top5}. 
We found these correspond to 
 \textsl{mastodon.social},  \textsl{pawoo.net},  \textsl{mastodon.xyz},  \textsl{octodon.social}, and  \textsl{mstdn.io}, in all the three cases.  
These are clearly among the most popular instances in Mastodon and, as expected, they allow their users to discuss a large variety of topics.         
Interestingly,   \textsl{mastodon.social} is always the top-1 instance regardless of the type of degree, whereas \textsl{pawoo.net} and \textsl{mastodon.xyz} alternate each other at the second and third rank.  
       							 					
Such instances are also well-recognized in the CCDF plot, where  
 we observe a probability of 50\% of having  at least 200 in-degree, which already  drops to 20\% for an in-degree around 600, and further decreases below 4\% for the outliers. 
The CCDF plot also displays the best fitting of power-law, lognormal, exponential and Poisson  distribution to the observed data. The    resulting fitting curves appear  to provide indications of lognormality and, to a limited extent, of power-law fitting.  

The above prompted us to  assess the corresponding statistical significance, whereby we resorted to a Kolmogorov-Smirnov test. Results are summarized in Table~\ref{tab:fitting-pl}.  
 In the first subtable, the high $p$-values 
suggest that the null hypothesis that the data are from a power-law distribution cannot be rejected, although this holds on a limited regime ($x_\textrm{min}$) starting from degree values, resp. in-degree and out-degree values, of the order of hundreds.  In particular,   for the in-degree case, note  that $x_\textrm{min}$ is above the mean of the distribution. 
  The remaining subtables in Table~\ref{tab:fitting-pl} correspond to four different scenarios we investigated for the lognormality fitting, namely (from top to bottom in the table)  full regime (i.e., whole observed data),   
removal of the outliers, removal of lower-degree  ($\leq50$) instances, removal of both outliers and lower-degree instances.  
As it can be noted,  the Kolmogorov-Smirnov test yielded high significance values   for the lognormality fitting in all cases, except when the outliers only are discarded; particularly, the significance is maximized when the lower-degree instances are removed (i.e., $p$-values from 0.687 to above 0.9). In such cases, the test informs us that  we cannot reject the null hypothesis and so we conclude that the observed data are lognormally distributed. 


\begin{table}[t!]
\centering
\caption{Power-law and lognormal fittings through \textit{Kolmogorov-Smirnov} test performed on the \textsc{Instances} network.}
\label{tab:fitting-pl}
\vspace{2mm}
\rmfamily
\begin{tabular}{|l|l||c|c|c|}
\hline
&  & \textit{degree} &  \textit{in-degree} &  \textit{out-degree} \\
\hline \hline
\multirow{3}{*}{power-law} &  $x_{min}$ & 890 & 588 & 457 \\
&  $\alpha$ & 2.987 & 3.166 & 3.057 \\
& $p$-value & 0.939 & 0.889 & 0.936 \\
\hline 
\hline
\multirow{12}{*}{lognormal} & interval & [1, 9\,062]  & [1, 4\,685]   & [1, 4\,377]   \\ 
& $\mu,\sigma$ &  5.54, 1.26  & 5.18, 1.24  &  5.11, 1.19  \\
& $p$-value & 0.113  &  0.187  & 0.113  \\
\cline{2-5}
  & interval & [1, 1\,307]   & [1, 878]   &  [1, 742]  \\ 
& $\mu,\sigma$ & 5.37, 1.15  & 5.05, 1.15   & 4.96,  1.09   \\
& $p$-value & 0.01  & 0.054   & 0.024   \\
\cline{2-5}
  & interval & [51,  9\,062]  &  [51,  4\,685]   &  [51, 4\,377]  \\ 
& $\mu,\sigma$ & 5.82, 0.95  & 5.53, 0.87   &  5.45, 0.81  \\
& $p$-value & 0.963  &  0.687  &  0.932  \\
\cline{2-5}
  & interval & [51, 1\,307]  &  [51, 878]  & [51, 742]   \\ 
& $\mu,\sigma$ & 5.65, 0.80  &  5.40, 0.74  & 5.31, 0.67   \\
& $p$-value & 0.47  &  0.646  &  0.816  \\
\hline
\end{tabular} 
\end{table}

\textbf{Sources and sinks.\ } 
We inspected the presence of instances having no incoming links (i.e., sources) as well as of instances having no outgoing links (i.e., sinks).  
 As reported in Table~\ref{tab:fullstats}, the percentage of both types of instances is not negligible, with the incidence of sources being nearly double than sinks. 
   This  might provide clues for the presence of users belonging to small instances (e.g.,  private ones) interested in   contents produced by users located in other instances; indeed, since small instances would host   few  users,  the lack of incoming links is plausible. On the opposite side, the percentage of sink instances sheds light on 
   that they might contain well-consolidated  user  groups, among which there is no need to interact with users belonging to other instances. This peculiarity might suggest sort of  self-sufficiency in the federation.

\textbf{Triadic closure.\ } 
We analyze how well the triadic closure principle is met in the \textsc{Instances} network, by looking at both transitivity (i.e., the probability that two incident edges are completed by a third one to form a triangle) and local clustering coefficient (i.e., how strongly connected are the  neighbors of a node). 

We observe a rather low value of transitivity (0.128), which is actually not surprising given the relatively low density of the network. By contrast, local clustering coefficient is  very high (0.836), and remains as such even when accounting for sink or source instances (0.687). 
This evidence is remarkable as it hints at    a federative structure among the instances.  
Note also that the dichotomy between a relatively lower transitivity and a higher    local clustering coefficient, and  more in general, the low correlation between the two statistics is typical of   networks characterized by a skewed degree   distribution.  
In this respect, the Mastodon \textsc{Instances} network also keeps this feature. 

As a further remark on length-2 closed loops (i.e., reciprocal edges), we observe a high fraction of reciprocal edges (above 65\%). As we shall further observe through our core decomposition analysis, reciprocity tends not to be limited to users within the same instance, but involves instances that can be placed very differently, from the periphery to the internal of the network, and vice versa.

\textbf{Degree assortativity.\ } 
One key structural property of a network at macroscopic level refers to degree correlation, or degree assortativity, which  measures how the probability of a link between two nodes  depends on their degrees~\cite{Newman2002,Newman2003}. 
Real-world social networks are often found to have positive degree assortativity, i.e., well-connected individuals are linked to other well-connected ones. 
A recent study has also shown that this evidence does actually hold for those social networks built upon shared memberships of group~\cite{Fisher2017}. 

Remarkably, the Mastodon \textsc{Instances} network  exhibits a degree assortativity which is significantly negative   (-0.253), which means that well-connected instances are connected to many  instances with few other connections. 
This     might be ascribed to the heterogeneous degrees characterizing the instances in the network. In this respect, we argue that, since instances can be bounded to specific topics, users belonging to different instances with diversified degrees  tend to interact with each other to reach a broader range of contents, thus ultimately improving their experience on the platform and increasing the speed of information transfer. 

The degree disassortativity, i.e., negative degree correlation, exhibited by the Mastodon \textsc{Instances} network outlines a novelty w.r.t. well-known centralized social networks. 
It should be noted that, unlike centralized social networks, Mastodon users' behavior is not impacted by recommendation mechanisms. Therefore, the followships tend to be built upon the topical interests and preferences that users have, which leads to a form of topically-induced link formation rather than a popularity-based attachment. Moreover, this further supports the strong interrelation between instances which, as we shall discuss in the next section, characterizes   peripheral and inner-core  locations  in Mastodon.

\subsection{Mesoscopic structural analysis}
\label{sec:mesoscopic}

We organize our presentation of the mesoscopic structural analysis of the \textsc{Instances} network into two parts: the first one is devoted to the evaluation of the community structure that is detected over the network, whereas the second part is concerned with the core decomposition of the network.


\begin{figure}[t!]
\centering
\includegraphics[width=0.85\textwidth]{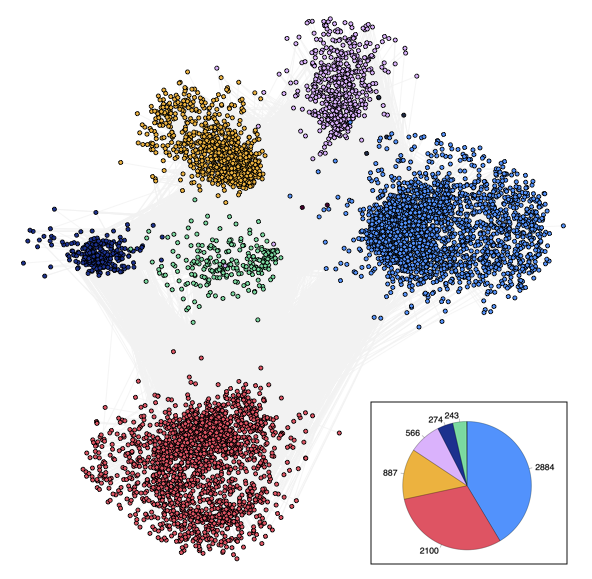}
\caption{Community structure on the \textsc{Instances} network obtained by the directed Louvain method. The displayed layout is based on the force-directed drawing Fruchterman-Reingold model, with weight 150 for edges incident nodes in the same community and weight 1 for edges incident nodes in different communities. The pie-chart in the bottom-right corner shows the size proportion of the communities, along with the size values for the top six largest communities.}
\label{fig:directed-louvain}
\end{figure}

\subsubsection{Community Detection} 
\label{sec:community detection}


We resorted to two well-known community detection algorithms for discovering communities in the Mastodon network of instances, i.e., the Louvain method~\cite{Blondel2008} and the Infomap method~\cite{RosvallB08}. 
 Louvain is a two-step, hierarchical greedy optimization method that attempts to maximize the modularity of a partition of the network, whereas Infomap optimizes the Map equation, which exploits the information-theoretic duality between finding community structure in networks and minimizing the description length of a random walker's movements on a network. 
It should be noted that both methods have been used with success for networks of many different types and sizes, and today, they are  the most widely used methods for detecting communities in large networks.  
For the Louvain algorithm, we exploited both the original, undirected implementation as well as the directed variant.\footnote{https://github.com/nicolasdugue/DirectedLouvain}  In the latter case, we also took into account the edge weights. As regards Infomap, we exploited its  weighted directed implementation.\footnote{https://www.mapequation.org/infomap/}


The number of communities found by the aforementioned algorithms is shown in Table~\ref{tab:fullstats}, for all considered scenarios (cf. notes marked with $^*$ and $^{**}$ below the table). 
Please note that we report two values for each case: the one within parenthesis corresponding to the total number of communities while the first value refers to  the number of communities that contain at least ten instances. 
Furthermore, since the Louvain algorithm optimizes modularity, we also report the modularity values corresponding to the community structures discovered by Louvain;   
in this regard, we find   evidence of modular structure within the \textsc{Instances} network, with modularity from about 0.29 (undirected network) to 0.35 (weighted directed network).



The number of   communities produced by the Louvain method ranges from 5 for the undirected scenario,  to 8  when applying the weighted directed variant; in the latter case,  only 2 out of 8 communities contain  less than ten instances.   Infomap, conversely, appears to detect  a much higher total number of communities; however, by inspectioning them, we found out that a large majority are poorly significant as they consist of less than ten instances.  Indeed, as reported in the table,  we point out that the two methods actually behave the same in terms of number of relatively large communities (i.e., 6 communities produced by either  method considering weights and orientation of edges).


Figure~\ref{fig:directed-louvain} illustrates the communities detected by the directed Louvain method on the \textsc{Instances} network.   
As it can be observed, two main communities arise in terms of size (displayed at the bottom and on the right in the figure), which together contain  about 72\% of the instances in the network.  Other two communities (on the top in the figure) also stand out, as they  contain about 21\% of the instances.  
Remarkably, albeit less evident for the smallest communities, we can observe  high connectivity between all of them, which hints at the high interrelation between instances belonging to different communities.

\begin{table}[t!]
\centering
\caption{Unweighted and weighted \textit{conductance} scores for the community structures obtained by Louvain and Infomap methods on  the  \textsc{Instances} network. The first row refers to the average over all pairwise scores between communities, whereas the other rows refer to comparisons between the three largest communities detected by the methods. Weighted variants of conductance account for edges weight when calculating volumes and cuts.}
\label{tab:conductance-stats}
\vspace{2mm}
\rmfamily
\scalebox{0.9}{
\begin{tabular}{|l||c|c||c|c|}
\hline
& \multicolumn{2}{c||}{Louvain} & \multicolumn{2}{c|}{Infomap}\\
\cline{2-5}
& \textit{unweighted} & \textit{weighted} & \textit{unweighted} & \textit{weighted} \\
\hline 	\hline
\textsc{Instances} & 0.285  & 0.239 & 0.037 &  0.035 \\ 
Top-1 vs. Top-2 & 0.315 & 0.191 & 0.370 & 0.219 \\
Top-1 vs. Top-3 & 0.816 & 0.881 & 0.561 & 0.721 \\
Top-2 vs. Top-3 & 0.100 & 0.059 & 0.231 & 0.335 \\
\hline
\end{tabular} 
}
\end{table}

 %

We also delved into the community structure obtained via the Louvain algorithm to get more insights into the community boundaries. 
The largest community (i.e., the rightmost one in Figure~\ref{fig:directed-louvain}) contains the most relevant Mastodon instance in the Fediverse, namely \textsl{mastodon.social}, which represents the first instance born with the Mastodon project. Consequently, given its role as a reference point in Mastodon, the large constellation of instances observed around it is not surprising. 
 Within the same community, we spotted \textsl{mstdn.io} and \textsl{octodon.social}. The above   three   instances are not topically bounded and use English as the primary language, with \textsl{mstdn.io} also embracing French and \textsl{octodon.social} extending its range of languages to Japanese and Portuguese. 
 We point out that these instances might share the same community given their relevance and longevity (e.g., \textsl{mstdn.io}  is up since early 2017) in the Fediverse.
Moving our focus to the second-largest community in size  (i.e., the one at the bottom of the figure),  
 we distinguished two relevant instances, namely \textsl{pawoo.net} and \textsl{mstdn.jp}. Their co-existence within the same community 
 is justified by the fact that both discuss various topics   and share the official language (i.e., Japanese).
 %
 %
 Furthermore, in the third-largest community, 
  we located the remaining instance, i.e., \textsl{mastodon.xyz},  of the previously mentioned  top-5 largest ones in Mastodon. This is general-purpose and  uses English and French as   primary languages. 
As a final remark, we point out that all the top-5 instances reported in Section~\ref{sec:macroscopic} are established in the largest three communities discovered by the Louvain algorithm, and hence their relevance is further strengthened due to their central role within these communities. 

We replicated the same explorative analysis also for the community structure detected by Infomap. Remarkably, mostly interesting patterns are found again. Indeed, the largest community still includes the same instances as in the case of Louvain, with the addition of \textsl{mastodon.xyz}, which in the Louvain solution is included in the third largest  community. Moreover, this similarity aspect also holds for the second largest community detected by Louvain and Infomap, respectively, which comprises the two Japanese instances. 
These coherent findings from two different community detection methods  would support an underlying logics in the organization of the instances and their interrelations.

\begin{figure}[t!]
\centering
\includegraphics[width=0.7\textwidth]{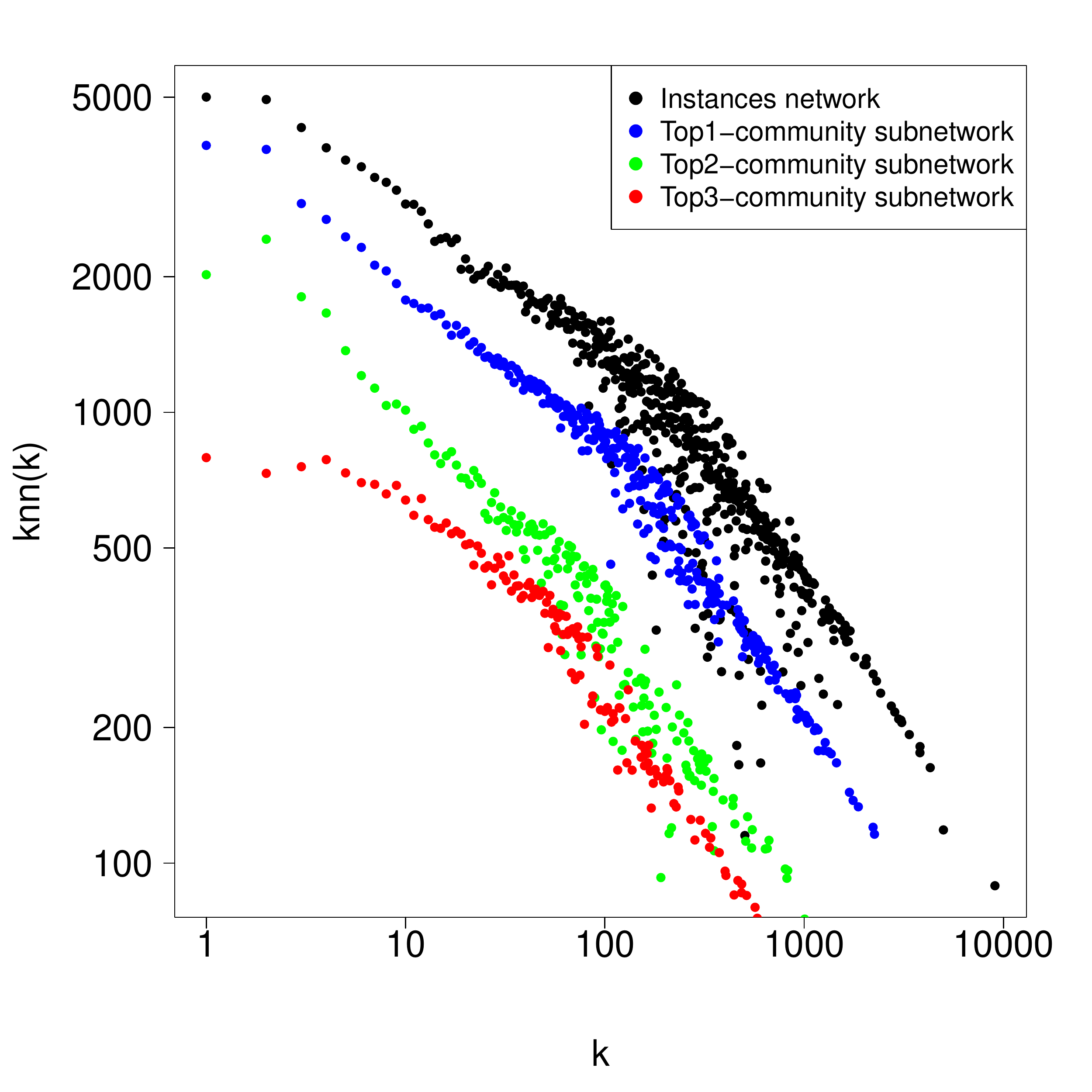}
\caption{Illustration of the average nearest neighbor  degree $knn(k)$ as a function of the degree $k$. Values on the x-axis and y-axis are log scaled.  The   three largest  community subnetworks are extracted from the structure obtained by the weighted directed Louvain method on the \textsc{Instances} network.}
\label{fig:knn-plot}
\end{figure}

We further investigated the community structures produced by Louvain and Infomap according to the aspect  of \textit{conductance}. This is defined as   the ratio between the cut size among two communities (i.e., the sum of the weights of the edges that link two sets of nodes) and the smaller of the volumes (i.e., the sum of the degrees of the nodes in a set) of the two communities; we considered  both the weighted and unweighted versions of conductance, where for the latter, the edge  weights   are equal to one.

As reported in Table~\ref{tab:conductance-stats}, conductance varies  depending on the community detection approach. 
Both methods induce good separation among communities in the \textsc{Instances} network, which is indicated by the low values of conductance, but in the Infomap solution  this is much more evident than in the Louvain community structure.  
Taking a  finer-grain perspective with a focus on the three largest   communities produced by the two methods, respectively, 
the analysis of the pairwise conductance among such communities yields two main outcomes: the conductance between the largest and the third largest communities ranges from about 0.56 (0.72, for the weighted version) w.r.t.  Infomap to above 0.8 when using Louvain; by contrast, regardless of the community detection method, both unweighted and weighted conductance is much lower for the pairs involving  the second largest community.  
We tend to ascribe these differences in conductance since, on the one hand, the largest communities have significant inter-community communication flow (and hence, high cut size) due to the involvement of  most relevant instances over the network; on the other hand,  this inter-community connection may be limited due to  different cultures and languages, as it is the case for the second largest community which indeed is centered around the Japanese-language  \textsl{pawoo.net} and \textsl{mstdn.jp} instances. 

The above remarks also prompted us   
to measure dependencies between degrees of neighbor nodes  in the   communities, by computing the \textit{average nearest neighbor  degree} distributions.  
%
  Figure~\ref{fig:knn-plot} reports the values of the average nearest neighbor  degree as a function of the degree $k$, which we denote by  $knn(k)$.  As it can be noted, the decreasing trend by $k$ is not only clear for the whole \textsc{Instances} network --- which is indeed aligned with the negative degree assortativity previously analyzed --- but also for the subnetworks induced from the top three communities    exhibiting disassortative traits. 


\subsubsection{Core Decomposition}  
\label{sec:coredecomposition}

The core decomposition of a network  graph consists in assigning each node with an integer number  (the core index)  capturing how well the node is  connected with respect to its neighbors. The result is  a  threshold-based hierarchical decomposition of the graph into nested subgraphs, based on a threshold ($k$) which is set on the degree of nodes~\cite{seidman_core}. 
The identification of such tightly-knit substructures, or cores,  has long been used for  understanding mesoscale structural characteristics of a network, with several applications related to the computation of the local importance of  nodes~\cite{MalliarosGPV20}, including the estimation of the spreading potential of nodes~\cite{Kitsak2010,CalioTB20}. A key advantage of core decomposition lays on theoretically grounded definition and uniqueness of its solution, which can also  be computed   efficiently  in linear time w.r.t. the number of edges in the input graph.   

Given a   graph $G=\langle V, E \rangle$ and  any subset $S \subset V$, let us  denote with $G[S]=\langle S, E[S] \rangle$ the subgraph 
of $G$ induced by $S$, where $E[S] = E \cap (S \times S)$.
 %
 For any choice of an integer value $k \geq 0$,  the $k$-core of a graph is the maximal induced  subgraph $G[C_k] = \langle C_k, E[C_k] \rangle$ such that the number of neighbors of every node $v$ in $C_k$ is at least $k$. 
 The  \textit{degeneracy} $K$ of the graph is the highest value of $k$ such that $C_k \neq \emptyset$. The  core associated with the graph degeneracy is also called the \textit{inner most core}. 
 The set of all $k$-cores (i.e., $V=C_0 \supseteq  C_1 \supseteq \dots \supseteq C_{K}$) represents the core decomposition of the graph.
Moreover, the \textit{core-index}, or \textit{coreness}, of a node  $v$ is the largest $k$ such that $v \in C_k$ and $v \notin C_{k+1}$. 
   Note also that the above  definitions originally apply to undirected graphs, however they are straightforwardly adapted to directed graphs so that the degree of a node may refer to either its in-degree or out-degree. 

\begin{figure}[t!]
\centering
\includegraphics[width=0.7\textwidth]{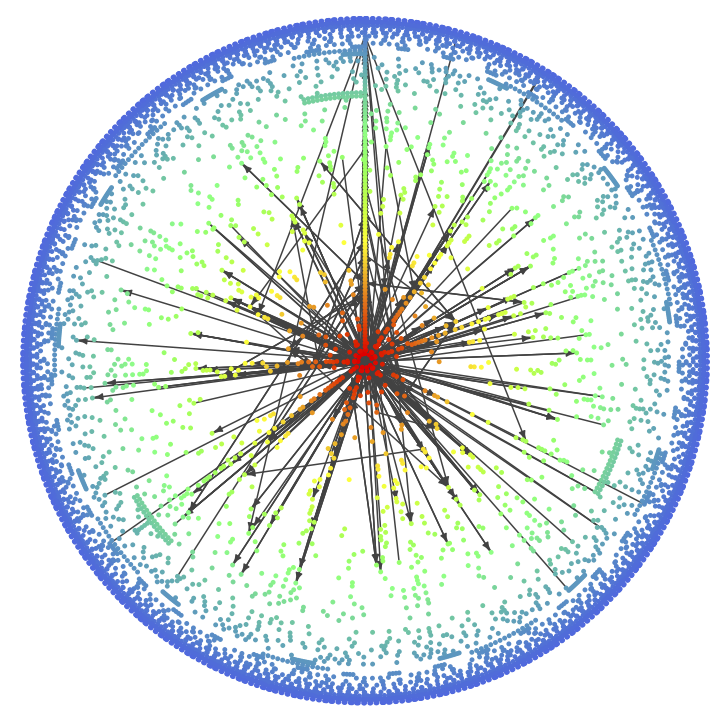}
\caption{Core decomposition of the \textsc{Instances} network,  based on node in-degrees.  Nodes having the same core-index are assigned the same color (inner-most,  resp.  outer-most core correspond to red,  resp.  blue).  To avoid cluttering, only edges having a weight greater  than the first quartile of (unique) edge weights are displayed.}
\label{fig:cores}
\end{figure}

The core decomposition of the \textsc{Instances} network revealed further hints at the presence of a federative mechanism. 
As reported in Table~\ref{tab:fullstats}, one major finding is the remarkable number of instances in the inner cores of the network, and even in the inner-most core, which ranges from 189 based on total degree to 208 based on in-degree, and also includes  the top-5 instances   previously discussed in Section~\ref{sec:macroscopic}. 
  This prompted us to explore the inner-most core of the \textsc{Instances} network, in its three variants, i.e., via a  total-degree-based, in-degree-based, and out-degree-based decomposition of the network. 
  
 A few remarks arise from the observation of the inner-most cores,  such as the high degeneracy under all the considered scenarios: indeed, we found out the values   201, 96, and 97 of degeneracy corresponding to a core decomposition based on total degree, in-degree, and out-degree, respectively.  
   As expected, density is high, and so is also the reciprocity --- note that statistics still refer to directed subgraphs, although the core-decomposition may have been generated based on the undirected (i.e., total degree-based) definition. 
 Although less evident, the disassortative trait is still present, denoting the coexistence in the inner-most core of instances exhibiting diversified yet   high degrees. 
 Given the high density, average path length and diameter almost halve  compared to their original values. 
 According to a more cohesive structure, transitivity   significantly increases  (nearly seven  times the value in the whole network) and aligns with the local clustering coefficient. 
 Also, the number of strongly connected components (i.e., \#SCCs) heavily shrinks to 1, resp. 2, for the indegree-based, resp. degree- and out-degree-based decomposition. 

\begin{figure}[t!]
\centering
\includegraphics[width=0.7\textwidth]{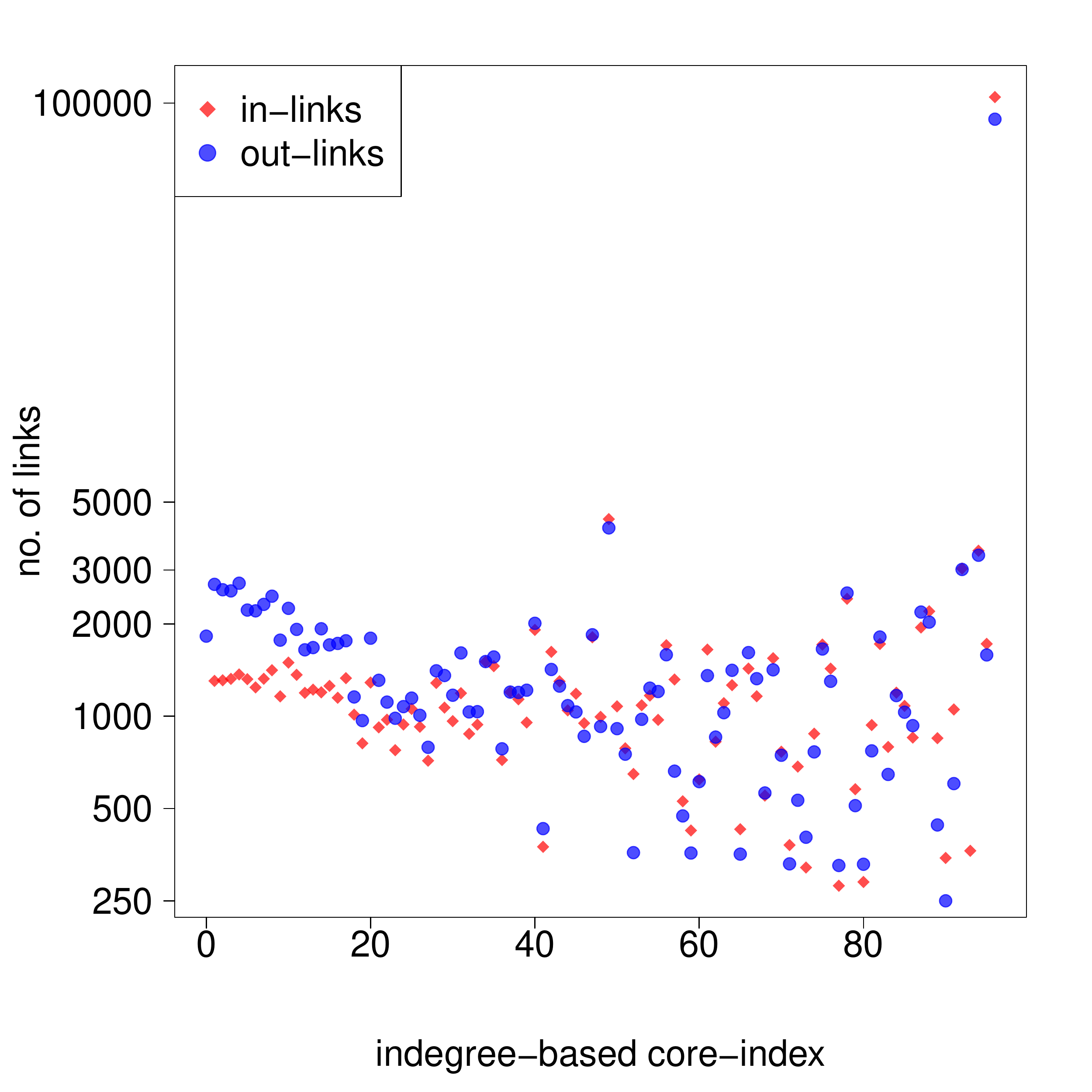}
\caption{Comparison between no.  of in-links and out-links for each indegree-based core-index of the \textit{instances} network.  Values on the y-axis are log scaled. }
\label{fig:scatter-cores}
\end{figure}

We also evaluated the presence of a community structure in the three variants of inner-most core. 
The (undirected) Louvain method  
determines an exceptional reduction in modularity compared to the value found for the whole network (i.e., from nearly 0.3 to 0.03-0.04), along with a small decrease in the number of significant communities (from 5 to 3). 
By contrast, when considering edge orientation and weights, we observe   a slight  decrease in modularity (from about 0.35 to 0.25) and    community structure size (from 8 to 4-6). 
Interestingly,   the number of communities detected by Infomap strongly  decreases  compared to the value observed for the whole network, with one giant community and the remaining 3-4  containing less than ten instances. 

Our major focus in this analysis was about whether there exists a significant amount of connections between the periphery and the inner parts of the \textsc{Instances} network. 
To this purpose, we started with a visual  exploration of the cores of the network, whose results are shown in Figure~\ref{fig:cores} for the indegree-based decomposition. 
Looking at the figure, the   display of the cores unveils evidence of directed links with a relatively high weight (i.e., solid lines in the chart), which  correspond to followships from  users of instances in the inner-most core to users of more peripheral instances, but also  followships in the opposite direction, even coming from the outer-most cores towards the inner-most core.  
This certainly relates to the negative degree assortativity previously emphasized, and it represents quite a novel pattern in social networks, which    usually  do not show direct links  between core and peripheral nodes. 

We further delved into this trait by examining the number of incoming links and outgoing links for each core-index, as depicted in Figure~\ref{fig:scatter-cores}. Lower  core-index values (up to about 20) correspond to a number of outgoing links that is significantly higher than  the number of incoming links, which is expected (i.e., instances within these cores tend to interact with more relevant ones, while the opposite   occurs less, or not at all). However, as the core-index value increases, unveiling an interesting feature of bidirectionality in the followships behavior for mid and internal cores. 
Further interestingly, some instances with high core-index (i.e., from about 70) show a prevalence of outgoing links.
 A final, remarkable trait of the \textsc{Instances} network is that a major portion of connections (i.e., about 190\,000, which is more than 87\% of the total edge set size) depart from and arrive to the inner-most core. 
 

\subsection{Discussion}

Our analysis of the structural characteristics of the  \textsc{Instances} network has revealed noteworthy features, which might set Mastodon apart from  well-known centralized social networks. Here, we summarize   such features, thus answering our third research question (\textbf{Q3}) on the ``fingerprint'' of the network of Mastodon instances. 

The evaluation at the macroscopic level has unveiled various unique traits of Mastodon. 
 Interesting aspects are already present  in the degree distribution of the instances:  indeed, as opposed to well-known centralized social networks that tends to fit a power-law distribution, Mastodon reveals a better and extensive fitting with a lognormal distribution, along with the presence of few    instances that show a degree up to one order of magnitude higher than the average degree in the network.  

Referring to the federation concept as a set of independent yet cooperating instances, we have found a number of aspects (e.g., high clustering coefficient values and percentage of reciprocal edges) indicating that Mastodon adopts a federative mechanism. We mark this mechanism as a \textit{mutual reinforcement} to reduce the sectorization bias that can characterize  the individual instances: indeed, they might be topically-bounded due to the decentralized nature of the platform, however, their  users generally look for a broader spectrum of topics, and hence they interact across different instances. We point out that, besides distinguishing Mastodon from other social networks, this trait is central to the platform itself, thanks to the shared protocol (i.e.,  ActivityPub) among instances.

Related to this mutual reinforcement is   the observed negative degree correlation (i.e., degree disassortativity), which represents another distinctive Mastodon feature. 
 This   indicates that users belonging to different instances with heterogeneous degrees tend to interact with each other, aiming at achieving a better user experience and increasing the speed of information transfer. 
 
The evaluation at the mesoscopic level has also  highlighted traits that contribute to determine the fingerprint of Mastodon. Through community detection, we shed light on the modular structures within instances, which provide further clues to the existence of a cohesive and federative framework among them. Further investigations have revealed how these modules might be composed and influenced. We observed topics, languages and temporal processes (e.g., their creation time) as main  influence factors.
 Moreover, through core decomposition, we spotted an unusual and conspicuous number of connections from the inner cores to the peripheral ones (Figure~\ref{fig:cores}), with a peculiar bidirectional balance between links observed starting from intermediate core-index values. Additionally, we observed that the majority of links between instances involve the inner-most core. 

In conclusion, we can state that  Mastodon reveals clear distinctive signs compared to what is commonly observed for other social networks, making it unique in multiple aspects. Among these traits, we detected the logical emergence of a federative mechanism, which  allows independent instances to cooperate with each other, thus connecting their respective users.


\section{Backbone of the \textsc{Instances} network}
\label{sec:backboning}
%


We focus here on a \textit{network simplification} task which is designed to ``prune'' the network graph, i.e.,   to detect and remove irrelevant or spurious edges  with the purpose   of making it easier unveiling hidden substructures of a network. 
One simple solution to the above problem is to exploit information on the edge weights so to remove all edges having weight below a pre-determined,  global threshold.  Unfortunately,  besides the difficulty of choosing a proper threshold for any input network,  this approach tends to remove all ties that are weak at network level,  thus discarding local properties at node level. 
 
By contrast,  a theoretically well-founded approach to filtering out noisy edges from a network  is based on \textit{generative null models}.  The general idea is to define a null model based on node distribution properties,  use it to compute a $p$-value for every edge (i.e.,  to determine the statistical significance of properties  assigned to edges from a given distribution),  and finally filter out all edges having $p$-value above a chosen significance level.  In other terms,  this allows to maintain only those edges that are least likely to have occurred due to random chance,  hereinafter referred to as \textit{backbone} of the network. 
Clearly, imposing lower significance level will yield to more restrictive substructures, thus giving place to a potential hierarchy of backbones. 

Identifying the backbone of a network graph allows us to isolate the latent structure of the network under analysis,  based on the removal of statistically insignificant edges.  In this regard,  statistical models for graph pruning have been conceived to deal with weighted networks,  so that the node degree and/or the node strength 
are used to generate a model that defines a random set of graphs resembling the observed network.     
One of the earliest methods is the \textit{disparity} filter~\cite{Serrano+2009},  which evaluates the strength and degree of each node locally.  The null hypothesis is that the strength of a node is redistributed uniformly at random over the node's incident edges. 
%
%
Unlike disparity, 
 the null model proposed by Dianati~\cite{Dianati2016} is maximum-entropy based and hence unbiased.  Upon it,  two models are defined:  the \textit{marginal likelihood filter} (\textit{MLF}), which is a linear-cost method that assigns a significance score to each edge based on the marginal distribution of edge weights, and the \textit{global likelihood filter}, which 
accounts for the correlations among edges.  While performing similarly,  the latter   is more costly than the former, 
 therefore we will consider the MLF model in our analysis. 


\subsection{Details on the disparity and MLF models for  weighted directed networks}
 
Both disparity and MLF were originally conceived for undirected networks, but they can easily be extended to weighted directed networks as well. 
 In the following, we provide a brief review of the two models, and refer the interested reader to the original works for further details~\cite{Serrano+2009,Dianati2016}.    

In the weighted directed network scenario, let us denote with $k_i^{in}$ and $k_i^{out}$ the in-degree and out-degree, respectively, for any node $v_i$. Also,  the total strength $s_i$ associated to the node  has two contributions, namely  the incoming strength $s_i^{in}$
 and the outgoing strength $s_i^{out}$, which are obtained by  
summing up all the weights of the incoming or outgoing links, 
respectively.

The disparity model aims to preserve the edges carrying a weight that represents a local significant deviation with respect to a statistical null model for the local assignment of weights by using the disparity function. Moreover, since in-degree and out-degree distributions are assumed not to be  necessarily correlated, incoming and outgoing links associated to a node  are  considered separately. 

To assess the effect of inhomogeneities in the weights at the local level, the following functions are defined for any node $v_i$: 
%
\begin{equation*} 
\Upsilon^{out}(v_i) = k_i^{out} \sum_j (p_{ij}^{out})^2, \\
\Upsilon^{in}(v_i) = k_i^{in} \sum_j (p_{ji}^{in})^2,
\end{equation*}
 
 \noindent 
 where the summation term in the first, resp. second, expression characterizes the level of local heterogeneity in the outgoing  edge weights, resp. incoming edge weights.  
  %
The null model for  the disparity filter requires  for each incoming, resp.  outgoing, link of $v_i$ with weight $w$, the calculation of the $p$-value   $\alpha_{ji}^{in}$, resp. $\alpha_{ij}^{out}$, that its normalized weight $p_{ji}^{in} = w/s_j^{in}$, resp. $p_{ij}^{out} = w/s_i^{out}$, is compatible with the null hypothesis. 
 The incoming and outgoing links that carry weights that  can be considered not compatible with a random distribution, can be filtered out  according to a chosen significance level $\alpha$; more specifically, all the incoming, resp. outgoing, links with $\alpha_{ji}^{in} < \alpha$, resp.  $\alpha_{ij}^{out} < \alpha$ reject the null hypothesis and hence can be considered as significant heterogeneities.

Let $T$ denote the number of unit edges, where each weighted edge is seen as multiple edges of unit weight. 
In the MLF model,  the null model assigns each edge to  nodes $v_i$ and $v_j$ 
 which are selected independently and randomly with probabilities proportional to their outgoing and incoming strengths, respectively. More in detail, the probability that an edge from $v_i$ to $v_j$ is associated to them is given by $p_{ij} = \frac{s_i^{out} s_j^{in}}{T^2}$. 
 The probability that a weight $w$ out of $T$ unit edges will choose nodes $v_i$ and $v_j$ as their endpoints is given by a binomial distribution with probability of success $w$ and number of trials $T$.  
   Therefore, the null model defines for every couple of nodes a probability  for their edge weight, $\sigma_{ij}$, depending on their strengths. 
    For each directed edge $(v_i,v_j)$ with weight $w_{ij}$, its $p$-value, denoted as $\pvalueij$, is computed as:
\begin{equation*}
\label{eq:pvalueDianati}
\pvalueij=\sum_{w\geq w_{ij}} \Pr[\sigma_{ij}=w | s_i^{out}, s_j^{in}, T] = \sum_{w\geq w_{ij}}  \binom{T}{w} p_{ij}^w (1-p_{ij})^{T-w}.
\end{equation*}
According to this null model, the higher the strengths of two nodes, the higher the weight of an edge connecting them is to be in order to be considered statistically significant. Conversely, the lower the strengths of the two   nodes, the lower the weight of a linking edge must be in order to be retained by the filter.

\begin{table}[t!]
\centering
\caption{Summary of structural characteristics of the \textsc{Instances} network before and after graph pruning. For each of the two graph pruning models, we tested under significance levels at 1\% and 5\%, which are indicated within brackets. }
\label{tab:backboning-stats}
\vspace{2mm}
\rmfamily
\scalebox{0.85}{
\begin{tabular}{|l||c|c|c|c|c|}
\hline
& \textit{unpruned} & \textit{MLF} & \textit{MLF} & \!\!\textit{Disparity}\!\! & \textit{Disparity}\\
&    & $(0.05)$   &   $(0.01)$  &   $(0.05)$  &   $(0.01)$ \\
\hline 	\hline
\#nodes & 6\,960 &  6\,454  & 6\,081 & 3\,949  & 3\,260 \\ 
\#edges & 216\,504  &  97\,141 & 68\,100 & 19\,208 & 12\,849 \\ 
reciprocity   &   65.1\% &  57.9\%  & 58.4\% &  67.8\% &  68.5\% \\ 
density   &   0.004  & 0.002  & 0.002 &  0.001 & 0.001 \\ 
average degree$^*$ &  41.966   &   21.390    &   15.860    &   6.429   & 5.182  \\
average in-degree   &  31.107  & 15.051 & 11.199 &  4.864 & 3.941 \\ 
\% sources &   12\% & 13.1\% & 14.1\% & 24.1\% & 26.5\% \\ 
\% sinks   & 6.6\%  &  6.5\%  & 6.7\%  & 6.9\%  & 7.9\% \\ 
degree assortativity$^*$ &   -0.274  &   -0.249    &   -0.224    &   -0.244  & -0.272 \\
degree assortativity & -0.253  & -0.231 & -0.214 & -0.253 & -0.283 \\
average path length  &   2.330  &  3.080  & 3.270 & 2.530  & 2.520 \\ 
diameter   &   5  &  6  & 7 & 5 & 6 \\ 
transitivity$^*$   &   0.128   &  0.070  &  0.066 &  0.024 & 0.018 \\ 
clustering coefficient$^*$  &  0.836  &  0.502  & 0.493 & 0.846 & 0.839 \\ 
clustering coefficient {\scriptsize \textit{(full averaging)}}$^*$ & 0.687 & 0.411   & 0.390 &  0.447 & 0.370 \\ 
\#strongly connected components     &   1\,305  &  1\,276  & 1\,278 & 1\,228  & 1\,121 \\ 
\#weakly connected components$^*$   &   1  &  1  & 1  &  1 & 1 \\ 
\hline \hline
modularity {\scriptsize by \textit{Louvain}}$^*$  & 0.289   & 0.495 & 0.540 & 0.469  & 0.495\\ 
\#communities {\scriptsize by \textit{Louvain}}$^*$ & 5 (5)  & 7 (7) & 7 (7)  & 8 (8) & 8 (9) \\

modularity {\scriptsize by \textit{Louvain}}$^{**}$  & {0.353}  & {0.442}  &  {0.442} & {0.369} & {0.375} \\ 
\#communities {\scriptsize by \textit{Louvain}}$^{**}$ & 6 (8) & 6 (8) & 7 (9) & 7 (9) & 7 (11) \\ 

\#communities {\scriptsize by \textit{Infomap}}$^{**}$ & 6 (54) & 10 (103) & 12 (82) & 6 (30) & 5 (38) \\ 
\hline
\end{tabular} 
}
\vspace{1mm}
{\footnotesize
\begin{flushleft} $^*$ Statistic calculated by discarding edge orientation 
\end{flushleft}
}
{\footnotesize
\begin{flushleft} $^{**}$ Statistic calculated by taking into account the edge weights 
\end{flushleft}
}
\end{table}

\subsection{Results on the pruned networks}
Table~\ref{tab:backboning-stats} reports on main structural characteristics of the \textsc{Instances} network after   graph pruning processes through the  application of  the MLF and disparity models (for the sake of presentation, values in the original, i.e., unpruned network are also reported). 

Using MLF,  55.1\% and 68.5\% of the edges are removed with significance level of $0.05$ and $0.01$, respectively. 
The pruning effect on the network size is much more evident when using the disparity model,  with at least 43.3\% of nodes and 91.1\% of edges removed.   
By contrast,    the reciprocity percentage in the MLF pruned networks is about 7\% less than in the original network, whereas the disparity pruned networks  show a small increase of  reciprocal edges (about 3\%). 
%

Interestingly, while it can be observed a significant decrease in the average in-degree --- from around 31 in the original, unpruned network up to 11-15, resp. about 4-5, 
 using   MLF and disparity,  respectively --- the degree assortativity
  remains negative in all cases, with a small decrease  in module under MLF, and comparable or increased in module for disparity with significance level $0.05$ and $0.01$, respectively.  
  Average path length  is also comparable, yet  with some fluctuations,  to its original value after applying disparity, while it marginally increases under MLF.  The latter determines an increase in the network diameter by one and two units for significance level $0.05$ and $0.01$, respectively, while disparity raises it by one unit only under $0.01$.  

We also observe a general reduction of the transitivity,  which becomes nearly one-half by MLF and even lower  by disparity model,  compared to its original value.   
The local clustering coefficient remains substantially unchanged   under disparity, while it is nearly halved   under MLF; however, when excluding source and sink nodes, the effect is similar for both pruning models and the gap from the original value is smaller. 


Both pruning approaches lead to a relatively small  decrease in   the number of strongly connected components,  which is slightly more evident under the disparity pruning.  
Looking at the community structures, the impact of the two pruning models varies  depending on the specific community detection method. As concerns Louvain, the number of communities has little variations, while the pruning is always beneficial according to an improved modularity. 
On the other hand, the number of communities produced  by Infomap almost doubles in the  MLF pruned network, which suggests that the pruning effect due to the MLF model would favor  the detection of  many small communities by Infomap. 



\subsection{Discussion}
To answer our fourth research question (\textbf{Q4}), we discussed how 
the backbone of the Mastodon network of instances is unveiled by leveraging   on graph pruning approaches based on  generative null models. 
These approaches allow  us to discard noisy edges in the network and shed light on its well-rooted underlying structures. 


MLF and disparity models showed to differ in pruning intensity, with the latter being more severe. To explain this, recall that the disparity filter  accounts for both the degree and strength of a node locally, whereas MLF relies on a maximum-entropy based, unbiased null model:  with this in mind, the disparity filter is likely to be more heavily conditioned on   in-degree and out-degree distributions that were observed in the \textsc{Instances} network  to best fit with a lognormal distribution and, to a less extent, with a power-law distribution, which is the best scenario of application of disparity~\cite{Serrano+2009}. 

A major remark that stands out is nonetheless a certain consistency with the structural properties of the  \textsc{Instances} network; that is, apart from a significant decrease in transitivity as an expected effect of the pruning, the observed structural characteristics in the pruned networks remain comparable to those of the original network, or even more emphasized as for the modularity. In particular,   the disassortative trait is still present in the backbone, which   certainly strengthens our previous finding  so to distinguish Mastodon from many other social networks in terms of negative degree correlation. 


It is also worth noticing that  the top-5 instances discussed in Section~\ref{sec:macroscopic}, regardless of the particular community detection method   
are still found after all the considered pruning scenarios,  
which indicates that these instances are important constituent of the backbone of  the \textsc{Instances} network. 
Overall, even the pruned scenarios exhibit traits related to a tight structure among instances. We recall that pruning approaches aim to filter out noise from networks, and therefore these traits are further strengthened following a cleaner perspective.

\section{Evolution of the network of Mastodon instances}
\label{sec:evolution}

In this section we answer our fifth  research question (\textbf{Q5}), namely ``\textit{how has Mastodon evolved during the last few years?}''. 
  To this aim, we organize our presentation into four parts: the first one is devoted to a comparison between the main network we introduced into this work, i.e.,  \textsc{Instances}, and the   network of instances  previously studied in~\cite{Zignani2018}; the second part focuses on the   subnetwork composed of online instances in our updated network; 
  the third part describes our analysis of  centrality of the instances based on the PageRank method;  
  finally, we report a summary of the main lessons learned from our study of the Mastodon growth.

\subsection{Comparison with the earlier Mastodon network}
\label{sec:evolution:earlier}

\begin{table}[t!]
\centering
\caption{Evolution of the \textsc{Instances} network through a comparison of structural characteristics with the earlier state-of-the-art network. Percentage changes refer to the increase/decrease of a statistic from the value observed in the earlier network. For the community statistics, percentage values refer to the number of communities containing at least ten instances. }
\label{tab:stats-growth}
\vspace{2mm}
\rmfamily
\scalebox{0.9}{
\begin{tabular}{|l||c|c|}
\hline
& \textit{Earlier}~\cite{Zignani2018} & \textit{\% change} \\
\hline 	\hline
\#nodes & 4\,015  				& +73\% \\ 
\#edges &   95\,221 			& +127\% \\
reciprocity   &   70.9\%   	& -8\% \\
density   &   0.006   			& -33\% \\
average degree$^*$ &    30.612    &  +37\%     \\ 
average in-degree   &  23.716  & +31\% \\
\% sources  &   5.63\%    		& +113\% \\
\% sinks   &  9.39\%  			& -30\% \\
degree assortativity$^*$ & -0.287  &    -5\%\\
degree assortativity & -0.291  		& -13\%\\
average  path length   &   2.340   	& 0\% \\
diameter   &   5   			& 0\% \\
transitivity$^*$   &   0.135   	& -5\% \\
clustering coefficient$^*$   &  0.848   	& -1\% \\
clustering coefficient {\scriptsize \textit{(full averaging)}}$^*$   &   {0.710}    &  {-3\%} \\ 
\#strongly connected components    &   604   &   +116\% \\
\#weakly connected components$^*$   &   1   &  0\% \\
\hline \hline 
modularity {\scriptsize by \textit{Louvain}}$^*$ &  0.356  & {-19\%}  \\ 
\#communities {\scriptsize by \textit{Louvain}}$^*$ &  4 (4)  & {+25\%}  \\ 

modularity {\scriptsize by \textit{Louvain}}$^{**}$ & {0.397} & {-11\%} \\
\#communities {\scriptsize by \textit{Louvain}}$^{**}$ &  3 (5) & {+100\%} \\

\#communities {\scriptsize by \textit{Infomap}}$^{**}$ & 5 (63) & {+20\%} \\
\hline \hline
degree-based degeneracy &   141 &  +43\%\\
degree-based inner-most-core {\#nodes} &   	120			&  +58\% \\ 
degree-based inner-most-core {\#edges} &   	11\,227			& +130\%  \\ 
in-degree-based degeneracy &  69  & +39\% \\
in-degree-based inner-most-core {\#nodes} &   	123			& +69\%  \\ 
in-degree-based inner-most-core {\#edges} &   	11\,401			& +152\%  \\ 
out-degree-based degeneracy &  70  & +39\% \\
out-degree-based inner-most-core {\#nodes} &   	115			&  +70\% \\ 
out-degree-based inner-most-core {\#edges} &   	10\,385			&  +155\% \\ 
\hline
\end{tabular}
}

\vspace{1mm}
{\footnotesize
\begin{flushleft} $^*$ Statistic calculated by discarding edge orientation 
\end{flushleft}
}
{\footnotesize
\begin{flushleft} $^{**}$ Statistic calculated by taking into account the edge weights 
\end{flushleft}
}
\end{table}



In this section we discuss a comparative evaluation  between the Mastodon instances network introduced in this work  and the earlier network presented in~\cite{Zignani2018}.    
By replicating on the latter the same structural analysis as we have   carried out on the \textsc{Instances} network, we gain insights into the evolution of Mastodon during the last 3 years, focusing on a set of  macroscopic and mesoscopic characteristics  that may confirm some traits or highlight novel trends.  
Table~\ref{tab:stats-growth} shows the structural properties that we observed on earlier Mastodon network along with the   percentage increase values obtained by the corresponding characteristics in our network (cf. Table~\ref{tab:fullstats}). 

Besides the already noticed larger size of the current network (with a +73\% of instances and a +127\% of links) compared to the earlier one, 
a general remark that arises is that the two networks of Mastodon instances show   differences in several characteristics, though with two major exceptions: the one relating to the average path length and the diameter, which are both unchanged, and the other one referring to both global and local clustering coefficient, which are slightly lower (i.e., from -1\% to 5\%) in the current network. This would hint at a consolidated small-world behavior by  the Mastodon instances. 

 As for the remaining properties, 
  we observe in our updated   network a small  decrease in the percentage of reciprocal edges,  
  while the percentage of sources and sinks is respectively  more than doubled and decreased of less than one third. 
  One reasonable explanation for that relates to  an increased presence of offline instances    in the \textsc{Instances} network whose further neighborhood however cannot be   explored via API. 
 It should however be noted that our updated  \textsc{Instances}  network still outperforms in size the earlier network even when removing their respective  sources and sinks, with a percentaage increase of 66\% in the number of instances.
  
 Also, in the \textsc{Instances} network, the number of strongly connected components is more than doubled, while the decrease in density (about one third) should be   related to a high increase in the number of sources, which is in turn an effect of our deep crawling.  
 
 A major aspect of interest we found out in the \textsc{Instances} network, that is, degree disassortativity, holds similarly for the earlier network as well. This is noteworthy as it unveils  consistency of this property in Mastodon over time. 
 


Our updated network also shows more  communities but a slighly less modular structure  than the earlier one. Again, this should be ascribed to the impact due to the increase in the number of source instances.  
%
As concerns core decomposition, we notice that a high degeneracy  also characterizes the earlier network, and the values according to three degree  variants   are actually close to those observed in the \textsc{Instances} network in proportion of the respective node-set sizes, with only 9\%, resp. 2\%, of change when considering degree-based decomposition, resp. in-degree- or out-degree-based decomposition.

\subsection{Narrowing the focus on online Mastodon instances}
\label{sec:evolution:online}

\begin{table}[t!]
\centering
\caption{Evolution of the \textsc{Instances} network through a comparison of structural characteristics with the \textsc{Online-Instances} network. Percentage changes refer to the increase/decrease of a statistic from the value observed in the \textsc{Instances} network. For the community statistics, percentage values refer to the number of communities containing at least ten instances. }
\label{tab:stats-growth-online}
\vspace{2mm}
\rmfamily
\scalebox{0.9}{
\begin{tabular}{|l||c|c|}
\hline
& \textsc{Online-Instances} & \textit{\% change} \\
\hline 	\hline
\#nodes & 1\,115  & -84\%  \\
\#edges &   75\,046	& -65\% \\
reciprocity   &   70.6\%   	& +8\% \\
density   &   0.06   			& +1400\%  \\
average degree$^*$ &   87.074   &  +107\%  \\
average in-degree   &  67.306  &  +116\%   \\
\% sources  &   2.1\%	& -83\%  \\
\% sinks   &  1.4\%	&  -79\%  \\
degree assortativity$^*$ & -0.290  &  +6\%  \\
degree assortativity &  -0.284 		& +12\%  \\
average  path length   &   2.020  	& -13\%  \\
diameter   &   4		&  -20\%  \\
transitivity$^*$   &   0.369 	& +188\%  \\
clustering coefficient$^*$   & 0.712 	& -15\%  \\
clustering coefficient {\scriptsize \textit{(full averaging)}}$^*$   &   {0.689}    &  0\% \\ 
\#strongly connected components    &   41   & -97\%  \\
\#weakly connected components$^*$   &   1   &  0\% \\
\hline \hline 
modularity {\scriptsize by \textit{Louvain}}$^*$ & 0.229 & -21\% \\
\#communities {\scriptsize by \textit{Louvain}}$^*$ &  5 (5)  & 0\% \\ 

modularity {\scriptsize by \textit{Louvain}}$^{**}$ &  {0.336} & -5\% \\
\#communities {\scriptsize by \textit{Louvain}}$^{**}$ &  4 (7) & -33\%  \\

\#communities {\scriptsize by \textit{Infomap}}$^{**}$ &  4 (15) & -33\%  \\
\hline \hline
degree-based degeneracy &   168 &  -16\%  \\
degree-based inner-most-core {\#nodes} &   	153			& -19\%  \\
degree-based inner-most-core {\#edges} &   	17\,449			& -32\% \\
in-degree-based degeneracy &  80  &  -17\%  \\
in-degree-based inner-most-core {\#nodes} &   	173			& -17\%  \\
in-degree-based inner-most-core {\#edges} &   	20\,320			& -29\%  \\
out-degree-based degeneracy &  82  & -15\%  \\
out-degree-based inner-most-core {\#nodes} &   	158			& -19\% \\
out-degree-based inner-most-core {\#edges} &   	17\,987			&  -32\%  \\
\hline
\end{tabular}
}

\vspace{1mm}
{\footnotesize
\begin{flushleft} $^*$ Statistic calculated by discarding edge orientation 
\end{flushleft}
}
{\footnotesize
\begin{flushleft} $^{**}$ Statistic calculated by taking into account the edge weights 
\end{flushleft}
}
\end{table}

Our further step towards a   comprehensive understanding  of the current landscape of  Mastodon instances is an analysis of the subnetwork corresponding to the online instances only, i.e.,  the \textsc{Online-Instances} network, which was previously introduced in Section~\ref{sec:models}. 
 The goal here is to check whether our findings on the overall \textsc{Instances} network reflect on its subnetwork of online instances as well, or on the contrary  there are significant deviations on some traits. 
 Table~\ref{tab:stats-growth-online} summarizes the comparison between the two networks.  

A first remark clearly arises from the reduction in both the number of instances (-84\%) and links (-65\%) compared to the \textsc{Instances} network. This is clearly expected due to the deep API-based crawling capabilities, yet at the same time it captures the currently active snapshot of Mastodon in the Fediverse. We conjecture that, after a few years of novelty and curiosity that Mastodon attracted, a steady-state has been reached as composed by those instances that might be recognized as most established for the Mastodon users.   
The subnetwork of  online instances also turns out to be more tightly knit than the \textsc{Instances} network, as indicated by the increase in reciprocity (+8\%),  density  (one order of magnitude higher), increase in average degree and in-degree (more than doubled),  decrease in average path length and diameter (-13\% and -20\%, respectively), 
the almost tripled transitivity,   halved number of strongly connected components, and decrease in the percentage of sources and sinks (about 80\%). 
The latter aspect suggests that the high percentages (particularly for sources) found in the \textsc{Instances} network is likely to be ascribed to the different status  (i.e., online or offline) of the instances, which affects   the search outcomes.  Nonetheless, since the \textsc{Online-Instances} network contains currently active and online instances, their users might be more engaged in  developing interrelations. 
  %
 %
Notably, the disassortative trait is still present in the \textsc{Online-Instances} network, thus confirming it as a  well-rooted feature in the network of Mastodon instances.



As concerns the community structure,  the \textsc{Online-Instances} network exhibits similar characteristics  to the overall \textsc{Instances}, with a relatively small  reduction in undirected modularity -- which becomes negligible for the weighted directed case --  and in the number of   communities   only  when accounting for edge weight and orientation.  Interestingly, we point out that the number of significant communities (i.e., those containing at least ten instances)  in the \textsc{Online-Instances} is the same for  both Louvain and Infomap method, which is consistent with the situation already observed in the \textsc{Instances} network. 
 Analogous considerations can be made for the core decomposition results: indeed, not only the degeneracy remains still high in all three variants -- with a decrease of about 15\% w.r.t.   \textsc{Instances} -- but also the sizes of the corresponding inner-most cores are  proportionally higher than those in  \textsc{Instances} according to the respective node-set sizes, with 405\%, resp. 419 and 403\%, of increase when considering degree-based decomposition, resp. in-degree- and out-degree-based decomposition. 
 This is remarkable as it provides evidence of a well-established trend in Mastodon as to have a conspicuous concentration of instances in the inner-most central region of the network.

\subsection{Instance centrality}

To further investigate the growth   of Mastodon, we evaluated how the \textit{importance} of the  instances developed during the years. 
To this aim, we resorted to the well-known \textit{PageRank} method~\cite{PR}, which assigns prestige scores to the nodes in a network.  Using this tool, we comparatively evaluated results obtained on all the networks derived from our crawling data,  i.e., the \textsc{Expanded-Instances}, the \textsc{Instances}, and the \textsc{Online-Instances} network, as well as the earlier state-of-the-art network. 
This approach allowed us to assess whether and to what extent the centrality of instances was affected by the introduction of new instances and/or the dismission of instances (e.g., instances went to offline mode for a relatively long period, cf. Section~\ref{sec:data}).

We assessed the strength of relatedness  between the   PageRank solutions obtained on the above networks   by means of two standard rank correlation methods, namely \textit{Kendall correlation coefficient}~\cite{abdi07} and \textit{Fagin's intersection metric}~\cite{FaginKS03}. 
The Kendall correlation $\tau$ evaluates the similarity between rankings represented through set of ordered pairs assigned to the same set of nodes, relying on the number of inversions of pairs needed to transform one ranking into the other. 
Given two rankings $\mathcal{R'}$ and $\mathcal{R''}$ obtained on the same set of $N$ items,  
 this non-parametric correlation is formally expressed as follows:
\begin{equation*} 
\tau(\mathcal{R'}, \mathcal{R''}) = 1 - \frac{2   \Delta   (\mathcal{P}(\mathcal{R'}),      \mathcal{P}(\mathcal{R''})     )}{N(N-1)},
\end{equation*}
where $\Delta   (\mathcal{P}(\mathcal{R'}),      \mathcal{P}(\mathcal{R''})     )$ denotes the symmetric difference between $\mathcal{R'}$ and $\mathcal{R''}$, i.e., the number of unshared pairs between the two rankings. 
The values returned by $\tau$ are within the range [-1, 1], where 1 indicates that the two rankings are identical   and -1 means that one rank is the reverse order of the other. 
 
The Kendall correlation 
assigns the same importance to all items, regardless of their position.
Therefore, we also employed Fagin's intersection metric $F$, which considers partial rankings 
and assigns higher weights to items at the top of the lists. 
Given two ranking lists $\mathcal{R'}$ and $\mathcal{R''}$, Fagin's metric is   formalized as follows: 
\begin{equation*} 
F(\mathcal{R'}, \mathcal{R''}, k) = \frac{1}{k}  \sum_{q=1}^{k}   \frac{|  \mathcal{R'}_{:q}    \cap  \mathcal{R''}_{:q}   | }{q},  
\end{equation*}
where       $k$ is a parameter to determine the number of items to consider from the top of both rankings, and $\mathcal{R}_{:q}$ indicates the set of nodes from the 1\textit{st} to the q\textit{th} position in the ranking. Hence, we refer to $F$ as the average over the sum of the weighted overlaps calculated considering the first $k$ nodes in both rankings.  Its values are within the range [0, 1], where the highest the value, the better the score.

The results of our evaluations are reported in Table~\ref{tab:ranking-analysis}.
Since Kendall correlation requires a comparison of rankings of nodes from the same set, the results obtained on  any pair of networks correspond to the shared set of nodes. 
 Note also that in all cases the  Kendall correlation values are high, and associated with 
  $p$-values equal to zero, which means  rejection of the hypothesis of absence of correlation. 
  

Let us first consider  the comparison between \textsc{Expanded-Instances} and  \textsc{Instances} networks (i.e., first column of Table~\ref{tab:ranking-analysis}).   The   Fagin's intersection values observed for various $k$ values 
 are very high, as expected. Indeed, as described in Section~\ref{sec:data}, non-Mastodon instances represent the boundary of our \textsc{Expanded-Instances} network, and we have only partial knowledge of them. Hence, the more central role taken by the Mastodon instances is reasonable. However, while these remarkably high values denote that non-Mastodon instances only slightly influence the Mastodon ones, the relatively smaller value obtained for $F_{10}$ indicates that this influence actually affects the PageRank scores of the  top-10 instances. 

The comparison between \textsc{Online-Instances} and  \textsc{Instances} (i.e., second column of the table) allows us to  inspect the role of online vs. non-online instances.  
Paying attention to $F_{10}$, the noticeable value indicates how the most central Mastodon instances are well-rooted in their role, regardless of potential variations, even evident, in the number of the online ones at any given time. Although a bit less evident, 
 this trend remains valid in the range [50, 100], while we observed a further decrease with $F_{1\,000}$; the latter might indicate the existence of relevant instances among the offline or temporarily inactive ones.

As concerns \textsc{Expanded-Instances} vs. \textsc{Online-Instances}, we observe in general slightly lower values w.r.t. the previous comparison, which  might be ascribed to  the absence of some offline yet relevant instances as well as to  the lack of contributions given by non-Mastodon instances.  However, the still high correlation values observed further support  the centrality of the currently online Mastodon instances.

The last comparison we consider is between the \textsc{Instances} and the \textit{Earlier} network.  
It is not surprising to observe a general decrease in the correlation values, given the temporal difference  between the two networks. Nonetheless, the corresponding ranking lists over the shared set of instances show a good Kendall correlation (above 0.6); in addition, the Fagin's intersection metric values are also quite high, for various $k$ values. Overall, despite the introduction of new instances after about three years, the results    indicate that a significant portion  of  those  instances that are shared between the two networks
have not drastically changed their strategic location in their respective networks. 
 

%

\begin{table}[t!]
\centering
\caption{Ranking analysis performed via Kendall's tau ($\tau$) and Fagin's intersection metric ($F$) with various $k$ values (indicated as subscripts) on the  PageRank solutions obtained on the Mastodon networks. 
}
\label{tab:ranking-analysis}
\vspace{2mm}
\rmfamily
\scalebox{0.8}{
\begin{tabular}{|l||c|c|c|c|}
\hline
&  \textsc{Expanded-Instances}    & \textsc{Online-Instances}    & \textsc{Expanded-Instances} & \textsc{Instances} \\
& vs. \textsc{Instances} & vs. \textsc{Instances}  & vs. \textsc{Online-Instances} & vs.\textit{Earlier}\\
\hline 	\hline
$\tau$ & 0.953 & 0.899 & 0.883 & 0.626 \\
\hline
$F_{10}$ & 0.860 & 0.935 & 0.886 & 0.569 \\
$F_{50}$ & 0.911 & 0.814 & 0.813 & 0.563 \\
$F_{100}$ & 0.918 & 0.832 & 0.835 & 0.545 \\
$F_{500}$ & 0.938 & 0.821 & 0.828 & 0.487 \\
$F_{1\,000}$ & 0.915 & 0.748 & 0.738 & 0.473 \\
\hline
\end{tabular}
}
\end{table}

\subsection{Discussion}

To investigate the evolution of Mastodon from an instance perspective, we compared the various networks presented in Section~\ref{sec:data}, keeping our  focus on the \textsc{Instances} network.
First, we performed a   comparison with the \textit{Earlier} network, which refers to about three years ago, as previously discussed. This   allowed us to highlight the structural development of Mastodon. Although the different sizes have some effect on the  measures, we noticed some well-established features. 
 In particular, we report the same average path length and diameter in both networks, whose consolidation suggests that Mastodon instances act in a small-world fashion. 
The \textit{Earlier} network also yields the same disassortative trait spotted for the more recent \textsc{Instances} network. Thus, we can establish the existence of this property over time. This consolidation allows us to consider the negative degree correlation as a distinctive trait of Mastodon w.r.t.  centralized social networks and paves the way for further studies   concerning interactions among instances through the federative approach. 
From a mesoscopic perspective, the evaluation of the core decomposition brings us  to confirm another important trait. Indeed, as initially shown by the \textit{Earlier} network and subsequently confirmed by the \textsc{Instances} one, Mastodon instances exhibit high degeneracy values, comparable when accounting for the different sizes.

Along with the temporal development, we analyzed the active status of Mastodon, focusing on the subnetwork induced from its online instances (i.e., the \textsc{Online-Instances} network). This narrowing produces an exceptional reduction in size ({-84}\% instances and -65\% links among them), which might indicate   a stationary state in the number of online Mastodon instances. Indeed,  the spotted online ones are found to be no longer transient (e.g., testing instances) or driven by the feeling of novelty,  but they are rather    well-defined and established.  Remarkably, the \textsc{Online-Instances} network still exhibits the disassortative trait. 
The core decomposition of the \textsc{Online-Instances} network  confirms the high degeneracy found in the other Mastodon networks, setting it as a well-established trait of Mastodon, along with a remarkable concentration of instances in the inner-most core. 

Finally, the study of the instance centrality based on PageRank allowed us 
 to assert that the most prestigious instances are settled in their roles, 
also considering the temporal development, 
 which indicates an actual evolution of the platform towards a stable state. The  instance-rankings computed over different pairs of instance networks have generally shown good or very high  correlation, with relatively small fluctuations mainly due to the centrality contributions from offline or boundary instances,   but also as an effect of the growth of the platform over time. 

Overall, and in light of all the above findings, we can state that Mastodon has reached sort of structural stability.  This well-established structure couples with a solid federative mechanism among instances, which acts as mutual-reinforcement towards the sectorization bias induced by the decentralized scenario. 

 
%
%

\section{Conclusion and Future Work}
\label{sec:conclusion}

Nowadays, DOSNs express an alternative, user-centric 
 approach to support online interactions among users, with attention on self-hosted networking services,  and ownerships  in terms of code of conduct   and moderation policies for each operating server. 
 Mastodon represents the most widely adopted and recognized platform in the Fediverse of DOSNs. 
 In this work, we provided the first in-depth analysis of the network of Mastodon instances based on the largest and most up-to-date Mastodon relational data existing so far, which was originally built in this work. 
  We analyzed our instance network models from different perspectives (i.e., macroscopic, mesoscopic, backbone) to highlight the main structural characteristics that describe Mastodon. 
Among these, we spotted remarkable traits that define the fingerprint of Mastodon, setting it apart from well-known OSNs. Finally, we investigated the evolution of Mastodon, by identifying   features that have changed over time and those representing supporting pillars.

Our next research stage will be focused on the underlying network of Mastodon \textit{users}. As discussed in this work, our collected data are fine-grained at a user level, although the stored information concern user memberships and relations only, i.e., social contents are discarded. 
In this regard, by still leveraging the network of links formed by users, we plan to study the behavior of users not only within their home instances, but also their relations across instances. 
In particular, we believe it would be interesting to investigate the roles that users may take in their instances, at various levels of activity, and possibly alternate or opposite roles that the same users may exhibit when interacting with users of other instances. This would allow us to also analyze the  dichotomy between contribution and consumption of information induced by the user behavior relations  over the instances of Mastodon and other DOSNs in the Fediverse.



\newpage
\section*{Appendix}

\begin{figure}[h!]
\centering
\includegraphics[width=0.6\textwidth]
{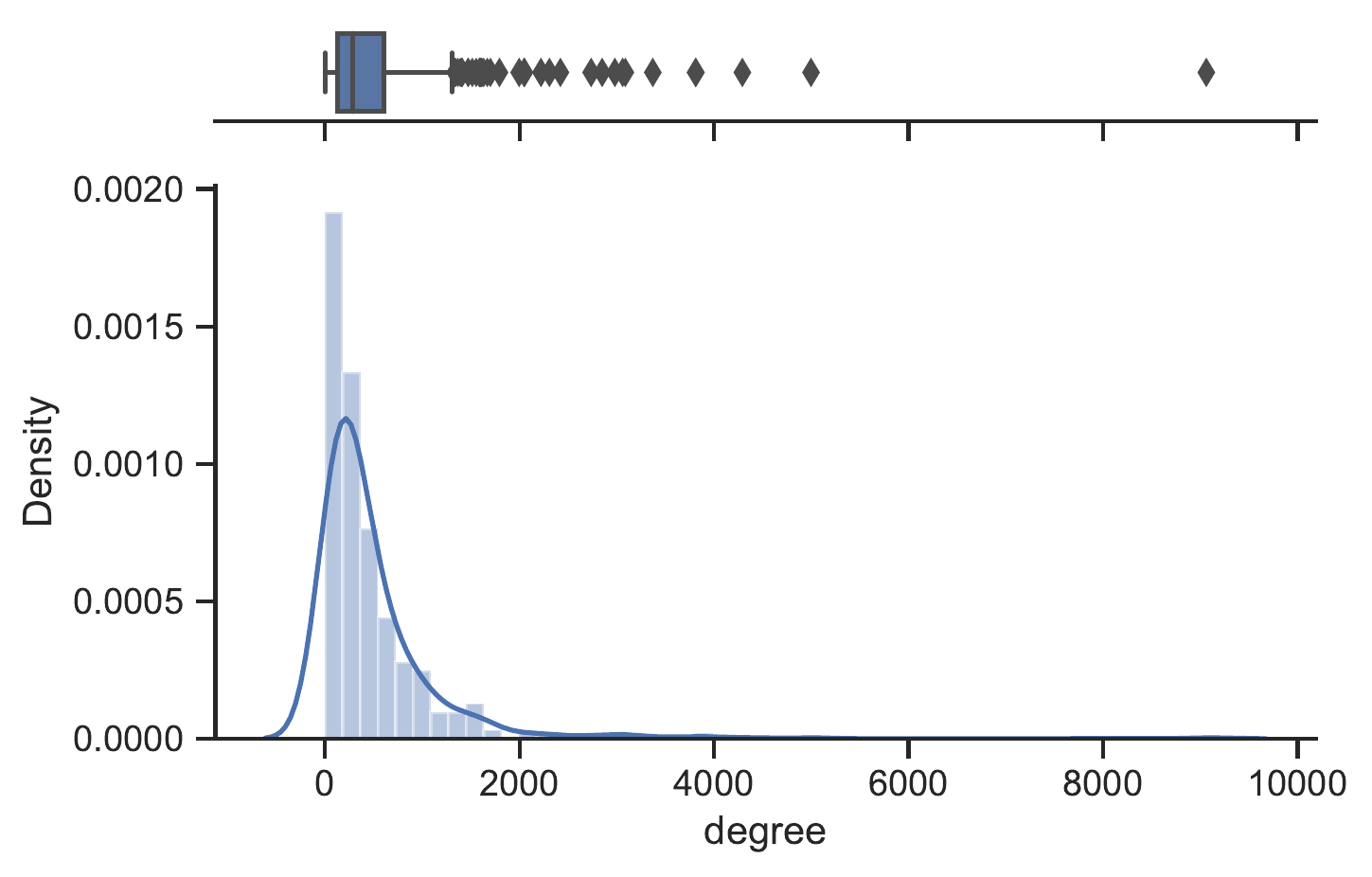} 
\\
\includegraphics[width=0.57\textwidth]
{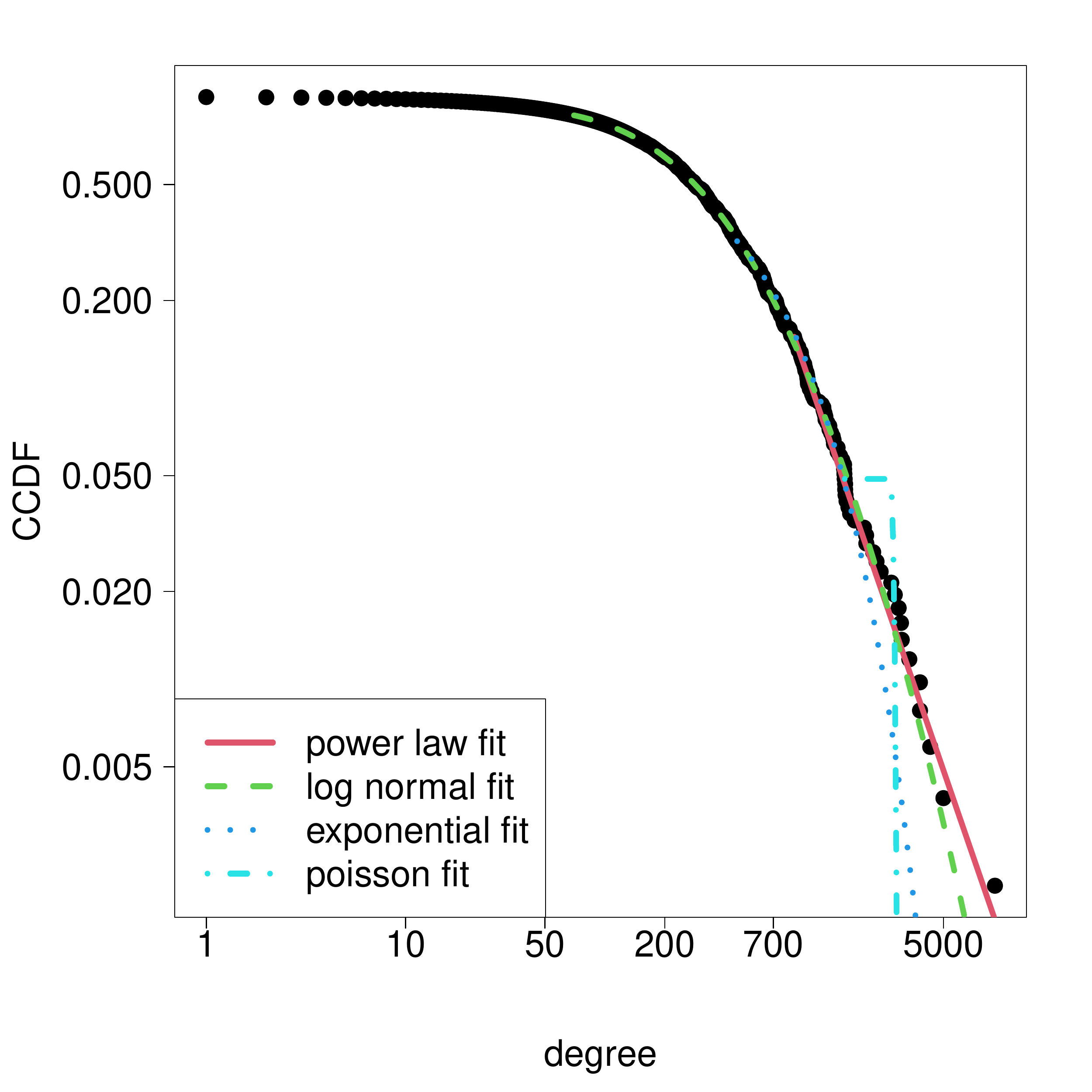} 

\caption{\textsc{Instances}   full-degree distribution: boxplot and Probability Density Function  (top), and Complementary Cumulative Distribution Function, with various distribution fittings (bottom).}
\label{fig:mastodon-instances-fitting-all}
\end{figure}

\end{document}